\documentclass[superscriptaddress,twocolumn]{revtex4}
\pdfoutput=1
\usepackage{graphicx,color,url}
\definecolor{b}{rgb}{0,0,1.0}
\definecolor{r}{rgb}{1,0,0}
\definecolor{g}{rgb}{0,1,0}

\begin{document}

\newcommand{\SZFKI}{
Institute for Solid State Physics and Optics, Wigner Research Centre for Physics,
Hungarian Academy of Sciences, P.O. Box 49, H-1525 Budapest, Hungary}

\newcommand{\MDB}{
Institute of Experimental Physics, Otto von Guericke University,
Universit\"atsplatz 2, 39106 Magdeburg, Germany. Fax: +49 391 67 18169; Tel: +49 391 67 18108; E-mail: ralf.stannarius@ovgu.de}

\title{Granular materials composed of shape-anisotropic grains}

\author{Tam\'as B\"orzs\"onyi}

\email{borzsonyi.tamas@wigner.mta.hu}
\affiliation{\SZFKI}
\author{Ralf Stannarius}
\affiliation{\MDB}

\begin{abstract}

Granulate physics has made considerable progress during the past decades in the understanding of
static and dynamic properties of large ensembles of interacting macroscopic particles, including
the modeling of phenomena like jamming, segregation and pattern formation, the development of related industrial applications or traffic flow control. The specific properties of systems composed of shape-anisotropic (elongated or flattened)
particles have attracted increasing interest in recent years. Orientational order and self-organization are
among the characteristic phenomena that add to the special features of granular matter of spherical or irregular particles. An overview of this research field is given.

\vspace{0.5cm}
\hspace{-2.2cm}
\noindent Final version published in Soft Matter, DOI:10.1039/C3SM50298H (2013)

\end{abstract}
\maketitle


\section{Introduction}

\label{sec:intro}

\subsection{Granular physics}

Granular matter, although being ubiquitous in everyday life, represents a class of materials that still fascinates us with startling, even counterintuitive behavior. The poet William Blake inspiringly described how to see a world in a grain of sand. Scientists have been equally fascinated to see a splendor of unexpected phenomena, and unique static and dynamic properties in ensembles of thousands and thousands of sand grains.
Interest in granulate physics dates back to the 18$^{\rm th}$ century when Coulomb \cite{Coulomb1773} in an essay considered the static angle of repose for a granular pile. Not only are the static properties of such a pile, like the force distributions inside \cite{Liu1995}, of scientific interest, but also its dynamics. When the grains on the slope of a pile begin to flow, one can observe avalanche dynamics and self-organized criticality \cite{Bak1987,Bak1988}. Grains that flow under shear tend to decrease their packing density, a phenomenon first described more than a century ago by Reynolds \cite{Reynolds1885}. Thus, sandpile physics is still a topic of intense scientific research today. Important aspects of granular physics are related to a variety of different topics, such as pedestrian \cite{Helbing2001} and car traffic \cite{Chowdhury2000,Nagatani2002,Helbing2005}, where cellular automata \cite{Biham1992,Nagel1992} compete with hydrodynamic models, pattern formation in granulates \cite{Umbanhowar1996,Jaeger1996,Aranson2006}, sand transport on geophysical scales\cite{Werner1995}, and on other celestial bodies \cite{Sagan1975,Lorenz2006}. Recently, an astonishing application of granulates in a robotic gripper was demonstrated \cite{Brown2010}. For a comprehensive list of references on granular physics, see the review by Kakalios \cite{Kakalios2005}.

Most experiments and theoretical studies were performed on spherical or irregularly shaped grains. However, grain shape may have significant qualitative influence on phenomena observed in granulates \cite{Frette1996}.
Particle shape can be described by several parameters such as aspect ratios, angularity, or convexity \cite{CEGEO}.
The grain shapes of sediments, soils, and rocks affect the packing of the material and the resulting pore geometries. They influence important transport properties, such as electrical conductivity, dielectric permittivity, diffusion, thermal conductivity, and hydraulic conductivity \cite{Friedman2002}. The stress distribution in a sandpile of shape-anisotropic ('anisometric') particles is very different from that in a pile of spherical grains, owing particularly to orientation effects \cite{Zuriguel2007}.

\begin{figure}[htbp]
\centering
 \includegraphics[width=\columnwidth]{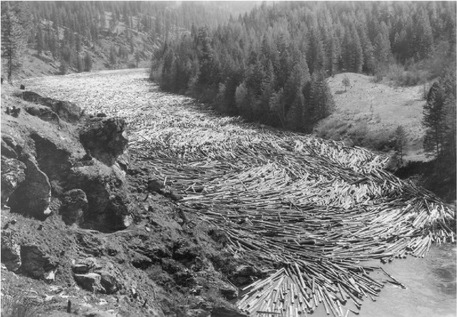}
 \caption{Log jam on Clearwater River, Idaho in 1931.
 The orientational order of the trunks includes local alignment, splayed and bent domains,
 preferential orientation at the lateral boundaries (banks) and a point disclination in the front.
 Photo courtesy of the Forest History Society, Durham, NC. [Image 5011]}
\label{fig:logjam}
\end{figure}

Various examples show that a careful characterization of packing, stresses, and flow of anisometric
grains can have immediate impact on the optimization of industrial processes and understanding complex natural phenomena. Modeling the formation of porous materials by depositing particles in gravity helps to understand the pore structure of sandstones (important for oil industry) or ceramics made for specific applications. Particles with even larger aspect ratio are the building blocks of numerous modern materials such as glass wool, filters, reticulate ceramics, or composite materials. In agriculture, storing and transportation of anisometric granulates is often not optimal due to a lack of accurate knowledge about the packing and flow properties of such materials. On an even larger scale, log jams are a typical consequence of the elongated shape of the individual floating trunks (Fig.~\ref{fig:logjam}).
In this review, we summarize recent progress in the understanding of orientational ordering, packing, and dynamics of granular materials composed of elongated or flattened particles.

\subsection{Characterization methods and modeling}

This section is not intended to give a exhaustive review of methodical aspects of the characterization and modeling of granular systems. Rather, we give a short overview of techniques that are referred to in the subsequent sections.
There are several common methods used to characterize grain ensembles, irrespective of grain shapes.
Modern experimental tools like X-ray Computed Tomography (X-ray CT) or Magnetic Resonance Imaging (MRI) are  increasingly used for characterizing not only coarse grain packing \cite{Nakagawa1993,Sakaie2008,Finger2006}, but also local packing structures and orientation at the grain level \cite{Zhang2006,Wegner2012,Borzsonyi2012}. Grain sizes accessible with X-ray CT equipment range from several micrometers \cite{Ketcham2001} to centimeters \cite{Wegner2012}. The method is very variable with respect to the choice of grain material. MRI, on the other hand, requires the presence of liquids containing NMR sensitive nuclei, the most abundant ones being protons. In order to employ this method for the analysis of granular beds, one is either restricted to NMR sensitive grains like seeds \cite{Nakagawa1993} or pharmaceutic pills, or one has to study solid particles immersed in a liquid \cite{Finger2006}. One can mark and trace particles with appropriate radio
 frequency and magnetic field gradient pulse sequences \cite{Ehrichs1995}. Spatial resolutions of MRI are comparable to those of X-ray CT. These non-invasive imaging techniques are highly superior to invasive characterization methods \cite{Montillet2001}.

The study of photoelastic materials in polarized light \cite{Howell1999, Daniels2008,Zuriguel2007} can be used to visualize force networks. However, like other optical techniques it is basically restricted to quasi-2D geometries.

Numerical approaches provide various data not easily accessible by experiments, e.g. contact numbers, and local stress anisotropy. Monte Carlo methods 
have proven useful, primarily for the determination of static structure parameters \cite{Buchalter1992,Buchalter1994,Stockely2003,Moradi2010}, but also for dynamic simulations \cite{Aspelmeier1998}. Well established numerical techniques to study structure and dynamics of granular ensembles are Discrete Element Methods \cite{Cundall1979} (DEM) (see, for example, Ref.~\cite{Zhu2007} and references cited therein).
DEM can treat the individual grains as either soft or hard particles.
The contact dynamics(CD) method  \cite{Moreau1994,Jean1999} considers impenetrable particles and dry friction between them. The equations of motion method are implicitly time integrated between contacts.
An event-driven version of molecular dynamics simulation that is particularly advantageous for very elongated  particles was proposed by Donev et al.~\cite{Donev2005}.

Random packings can be generated by molecular dynamics methods, but these techniques have some problems. Random sequential addition, or absorption, (RSA) \cite{Hinrichsen1986} has been employed in several studies, but the resulting final packing fractions not always seem to be realistic \cite{Williams2003}. The random close packing fraction is underestimated, particularly for large aspect ratios of the grains. Isotropy of packing is another issue.
A so-called mechanical contraction method (MCM) \cite{Williams2003}, and hybrid (MCM + DEM) approaches \cite{Wouterse2009} have been employed. In iterative steps, particle positions in a random dilute ensemble are  rescaled, subsequently overlapping particles are shifted into free volumes until the system is congested \cite{Williams2003}. Random dense sphere packings can be obtained using Voronoi tesselations \cite{Hinrichsen1990}, but this method seems to be impractical for shape-anisotropic particles.

\section{Static granular assemblies}
\subsection{Packing of shape-anisotropic grains}

\label{sec:packing}

Analyzing the packing of assemblies composed from grains with a specific shape provides information on the structure, symmetry, and macroscopic physical properties of the bulk. In particular, flow and jamming of granular materials are closely related to packing problems. A recent general review describes the current state of art for packing of spheres and various other shapes \cite{Torquato2010}. Here, we focus on simple anisometric (prolate or oblate) shapes.
The packing fraction of assemblies made of anisometric grains spans a wider range than for spherical
particles, which is reflected in a larger difference between the Random Close Packing (RCP) and Random Loose Packing (RLP). As we will see, a more compact RCP is related to the additional rotational configurations of   elongated particles, viz. compared to the case of spheres, more contacts are needed to ensure jamming. An additional factor for the possible increase in packing density is the orientational order of the particles. Analyzing the
spatial autocorrelation function (SAF) and introducing an effective particle size \cite{Matzler1997} are useful concepts to characterize packing.
In confined systems, an important factor for ordering is the effect of the bounding walls \cite{Zhang2006,Moradi2010}. Consequently the rich variety of packing-related phenomena has already attracted a lot of activities in {\it two-dimensional systems} (monolayers), where the role of the bounding walls is enhanced compared to three-dimensional systems.
For the experimental study of packing problems, two-dimensional (2D) systems offer some advantage over 3D systems, since they can be implemented in horizontal layers, so that in-plane effects of gravity can be disregarded.

\subsubsection{Two-dimensional systems}  \

The case of large aspect ratio (length/diameter $L/D>10$) rods was investigated experimentally and numerically (Monte Carlo) for a 2D pile \cite{Stockely2003}. The typical length of orientational correlations was found to be 2 particle lengths. To model shorter grains, two to seven spheres were welded together to form simple arrangements \cite{Rankenburg2001}. In these experiments, ordered domains formed (during annealing by external vibration) much more easily than in three dimensions. The rotational symmetry of the particle shape was identified as a useful predictor of how easily the grains order, and also a potential guide for identifying two-dimensional shapes that remain random after annealing.

Numerically, a system composed of frictionless ellipses shows fundamentally different behavior
near jamming compared to the case of circular disks \cite{Mailman2009,Zeravcic2009}.
Circular grains behave exceptionally: Packings are isostatic at the jamming point, all nontrivial vibrational modes increase quadratically with deformation amplitude.
The number of contact points per particle is 4, two times the number of degrees of freedom. Even packings of dimers (overlapping rigidly connected circular disks) are found to be isostatic, but those of ellipses with equal aspect ratio are not \cite{Schreck2010}. Packings of ellipses in general possess quartic modes, characterized by collective rotational motions. Interestingly, simulations show that already a 2\% elongation of the particle shapes changes the contact number to approximately 4.4 for ellipses and to 6 for dimers.
When the aspect ratio $L/D$ is varied, a sharp increase of the packing fraction versus  $L/D$ is found both for ellipses and dimers \cite{Shen2012}. However, one finds substantial differences between the packing fraction characteristics of dimers and ellipses for $L/D>1.5$ \cite{Shen2012}. In these simulations, bidisperse mixtures were used to avoid crystallization.

Often, modeling of non-spherical particle shapes can be facilitated when the grains are approximated
by chains of beads or disks. For this purpose, monomer, dimer and trimer systems of (non-overlapping) spherical particles were studied experimentally and theoretically in 2D \cite{Olson2002}. As expected, monomers order into a triangular lattice, while dimers show a novel quasi-ordered state in which only translational, but no bond-orientational order exists. Trimers (three spheres in a row) possess neither translational nor orientational order. Not only are the static distributions qualitatively different for the three species, but
the particle geometry has also a dramatic influence on the flow dynamics, which is discussed below in Sec.~\ref{sec:flowing}.

\subsubsection{Three-dimensional systems}  \

\begin{figure}[htbp]
\centering
 \includegraphics[width=6cm]{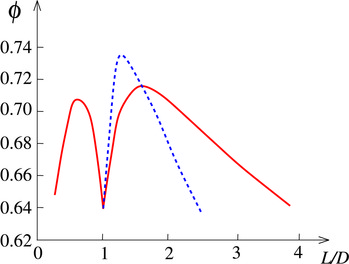}
\caption{Packing as a function of particle aspect ratio $L/D$ for axisymmetric ellipsoids \cite{Donev2004}
(continuous lines)
and asymmetric ellipsoids with axis ratios $a/b=b/c$ (dashed line) from numerical simulations.
Image sketched after Ref.~\cite{Donev2004}.}
\label{packing_aspect_ratio}
\end{figure}

In 3D, it has been demonstrated experimentally, as well as in numerical simulations, that random packing can be much more efficient for {not too large aspect ratio} ellipsoids than for spheres \cite{Donev2005}. As it is seen on Fig.~\ref{packing_aspect_ratio} (adapted from Ref.~\cite{Donev2004}), the packing fraction increases from $\approx 0.64$  for randomly packed spheres to 0.71 for axially symmetric ellipsoids with aspect ratio near 0.6 (oblate) and 1.5 (prolate). For completely asymmetric ellipsoids with axis ratios $a/b = b/c$, the random packing can almost reach the packing density of the face centered cubic lattice of spheres, $\approx 0.74$ \cite{Donev2004,Donev2005}.
More recently, simulations were extended to the deposition of general ellipsoidal particles with arbitrary axis ratios \cite{Baram2012}.
The rapid cusp-like increase observed near $L/D=1$ and $\phi\approx 0.64$ is attributable to the increasing number of contacts
due to the additional rotational configurations of the ellipsoids. More contacts, and consequently
higher packing densities, are needed to ensure jamming of ellipsoids, compared to spheres.
Some features of this packing characteristics can be reproduced with a simple one-dimensional model \cite{Chaikin2006}.
The above features have been observed for various sample preparation methods, including
numerical deposits with simple friction (no specific static friction) \cite{Baram2012},
a generalized form of the Lubachevsky-Stillinger algorithm \cite{Donev2007}, or even experimental
systems where ellipsoids were compacted in a centrifuge with several hundred $g$ \cite{Sacanna2007}.
A quantitative comparison of these systems, however, must be carried out with some reservation,
since not all methods produce randomly disordered ensembles. The influence of order in ellipsoid packings   deposited under the influence of gravity is detailed in the following Section
\ref{sec:pack_order}.
An isocounting conjecture, which states that for large disordered jammed packings
the average contact number per particle is twice the number of degrees of freedom per particle, does in general
not apply to nonspherical particles \cite{Donev2007,Chaikin2006}.

Figure~\ref{packing_aspect_ratio} also indicates decreasing packing fraction for particles with {\it large aspect ratio}. For the limiting case of infinitely thin rods, the random packing density goes to zero. Experimental studies provided accurate data for the dependence of the density in deposits of large aspect ratio rods made of wood, copper wire, or plastic \cite{Philipse1996,Philipse1996c,Novellani2000} on aspect ratios. These findings were later confirmed by numerical simulations \cite{Williams2003}. The experimental and numerical data are consistent with the suggestion based on a simple excluded volume argument \cite{Philipse1996,Philipse1996c}, namely that the packing fraction $\phi$ is related to the aspect ratio $L/D\gg 1$ and the average number of contacts experienced by a given particle, $\gamma$, as

\begin{equation}
\frac{\gamma}{\phi}=2\frac{L}{D}.
\label{eq:philipse}
\end{equation}

\begin{figure}[htbp]
\centering
 \includegraphics[width=7cm]{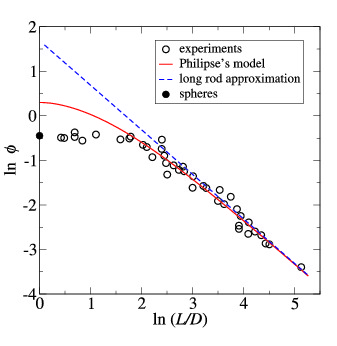}
 \caption{Logarithmic sketch of the measured packing fraction $\phi$ as a function of the aspect ratio $L/D$
 for (sphero-) cylinders \cite{Philipse1996,Philipse1996c,Milevski1978,Nardin1985} (circles), Philipse's model [Eq.~(6) in Ref.~\cite{Philipse1996}] (solid line) and the asymptotic approximation,  Eq.~(\ref{eq:philipse}), with $\gamma=10.8$ (dashed line). The figure is adapted from Ref.~ \cite{Philipse1996}. Note that our definition of the parameter $L$ differs from that of Philipse \cite{Philipse1996}, who defined $L$ as the bare cylinder length excluding the hemispheric caps.}
\label{fig:philipse}
\end{figure}

For the case of high aspect ratio sphero-cylinders, $\gamma$ was found to be around 10 in experiments \cite{Philipse1996,Philipse1996c,Blouwolff2006} and numerical simulations \cite{Wouterse2009}
(see Fig.~\ref{fig:philipse}).

\begin{figure}[htbp]
\centering
 \includegraphics[width=\columnwidth]{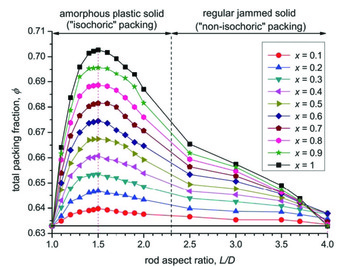}
 \caption{MCM computer simulation by Kyrylyuk et al. of the jamming of binary
mixtures of hard frictionless spheres and sphero-cylinders as a function of the
shape anisotropy $L/D$ for various mixture compositions.
$x=\phi_{\rm r}/\phi$ is the relative rod volume fraction, with the total packing volume $\phi=\phi_{\rm r}+\phi_{\rm s}$, indices $r$ and $s$ refer to rods (sphero-cylinders) and spheres, respectively.
Figure reprinted with permission from Ref.~\cite{Kyrylyuk2011b}.
}
\label{fig:kyrylyuk2011a}
\end{figure}

\begin{figure}[htbp]
\centering
 \includegraphics[width=\columnwidth]{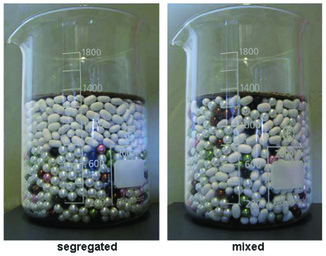}
 \caption{Spherical beads and sphero-cylindrical sweets (TicTacs with aspect ratio 1.72) in graded beakers.
 Isochoric ideality is evident: both segregated and mixed states have almost the same volumes.
 Random packing was achieved by gentle shaking and tapping.
Figure reprinted with permission from Ref.~\cite{Kyrylyuk2011b}.
}
\label{fig:kyrylyuk2011b}
\end{figure}

So far, ensembles of monodisperse particles with identical shapes have been considered.
Additionally, mixtures of ellipsoids with similar shapes and moderate size polydispersity
have been studied in numerical simulations \cite{Baram2012}. The influence of polydispersity
was found to be much lower than the influence of shape variations. A related aspect
is the mixture of grains with different shapes.
Recent numerical simulations and experiments showed that for mixtures of small aspect ratio
sphero-cylinders and spheres, an ideal mixing law can be found for the entire concentration
range \cite{Kyrylyuk2011b}: the packing fraction is linearly related to the mixing
ratio (Fig.~\ref{fig:kyrylyuk2011a}).
This means that the jammed system occupies the same volume independent of the particle dispersion,
in other words, irrespectively whether the system is segregated or perfectly mixed (Fig.~\ref{fig:kyrylyuk2011b}).
The aspect ratio dependence was the same for all compositions in the sense that the density maxima
(i.e. optimal random packing) corresponded to the same aspect ratio, regardless of the mixture composition.
For mixtures of sphero-cylinders with different aspect ratios, it was demonstrated that having a density maximum
with increasing aspect ratio is a universal feature of such ensembles  \cite{Kyrylyuk2011}.

\subsection{Effect of orientational ordering on packing}
\label{sec:pack_order}

Thus far we considered random packings, now we focus on how orientational ordering influences the packing. Such orientational order can occur as a consequence of various influences. It can spontaneously arise during the preparation of a system, e.g. pouring of particles into a container under the influence of
gravity. It can also be induced by container boundaries, or generated by shearing or shaking.

Even the preparation of random packings can be difficult for shape-anisotropic grains. An axially symmetric oblate ellipsoid, for example, has a tendency to adopt a vertical short axis orientation when dropped into a container. For prolate objects, the long axis tends to adopt a horizontal orientation.
Monte Carlo simulations of the packing of prolate or oblate ellipsoids poured into a 3D container
revealed such a preferential alignment in the gravitational field, which becomes stronger with increasing shape anisotropy \cite{Buchalter1994}. For sufficiently elongated particles, no translational order is present. All packings were found to be amorphous, but still exhibited long-range orientational order.
This is in sharp contrast with the behavior of a 2D system, where for small aspect ratios, the packings
were polycrystalline \cite{Buchalter1992}. In a two-dimensional system, packings become glassy only for
sufficiently large aspect ratios. The density of the packings is not a monotonic function of the aspect
ratio, and exhibits a global maximum for slightly elongated particles \cite{Buchalter1994}.

The presence of the {\it system boundaries} often has local effects on packing and ordering in boundary regions.
The packing of cylinders has been accurately studied by x-ray microtomography in a cylindrical container
\cite{Zhang2006}. At the highest overall packing densities, near-wall porosity becomes nearly equal to
interior porosity, and significant global ordering occurs near the wall.
Monte Carlo simulations of a confined hard ellipse fluid confirm these observations \cite{Moradi2010}.

Another interesting experimental configuration is the case of a {sandpile} that was already mentioned in the introduction. A peculiar feature of this
geometry is that the vertical stress measured below the pile has a minimum beneath the apex of the pile.
This problem has consequences in numerous practical situations, e.g. road and dam construction.
Zuriguel et al. \cite{Zuriguel2007} have shown that when the (2D) pile is made of ellipses, the dip beneath the pile was much more pronounced than for piles of circular particles. The amplification of this effect was related to orientational ordering of the ellipses.

Both the development of orientational order of anisotropic particles under shaking and compaction as a result of alignment are connected with an agitation of the material, and thus related to partial or complete fluidization of the granular bed. Compaction under vertical shaking has been described by various authors (Refs.~\cite{Villarruel2000,Lumay2004,Lumay2006,Vandewalle2007} and others). Numerical simulations by Donev et al.~\cite{Donev2004} showed that shearing of a random packing of oblate ellipsoids can introduce orientational order, which further increases the packing density. Shaking and shear experiments and their relation to orientational order will be described in detail in Secs.~\ref{sec:shaken} and \ref{sec:sheared}.

\subsection{Stress anisotropy}

\label{sec:stress}

The geometry of packing and the local stress distribution in a {\it vertical monolayer of deposited rods} have been studied experimentally and numerically \cite{CruzHidalgo2009,CruzHidalgo2010,Kanzaki2011}. For non-cohesive particles, the packing topology is dominated by the formation of ordered structures consisting of aligned rods. Elongated grains tend to align horizontally and the stress is transmitted mainly from top to bottom, revealing an asymmetric distribution of local stress \cite{CruzHidalgo2009,CruzHidalgo2010}. Kanzaki et al.  analyzed how this stress distribution changes if the two-dimensional container (silo) is partially discharged \cite{Kanzaki2011} through an orifice at the bottom. As we will see in section \ref{sec:flowing}, the flow pattern for elongated grains is different from that of circular ones, and it involves regions with strong velocity gradients (shear zones) where particles are reoriented by the flow. This leads to positional disorder and a reduction of the vertical stress propagation
during the formation of the deposit, prior to the partial discharge. With the introduction of cohesive forces, the preferred horizontal orientation disappears \cite{Hidalgo2012}. Very elongated grain shapes and strong attractive forces lead to extremely loose structures and diminished asymmetry in the distribution of local stress \cite{Hidalgo2012}.

{\it Two-dimensional packings} of elongated particles subject to biaxial shearing were studied in numerical simulations
\cite{Azema2010,Azema2012} using the contact dynamics (CD) method. The distribution of forces and the system's  dilatancy were studied as a function of the aspect ratio of the grains. The properties of the weak and strong force networks were analyzed in detail.  These studies revealed that in contrast to packings of disks, where the contacts in the weak network are on average perpendicular to the contacts in the strong network, the contacts for elongated particles are, on average, oriented along the major principal stress direction both in weak and strong networks. Other CD simulations of biaxial compression of disks or elongated polygonal shapes \cite{Nouguier-Lehon2003,Nouguier-Lehon2010} in 2D revealed that samples consisting of angular particles show larger shear strength and dilatancy due to face-to-face contacts that
restrict rotations. The friction angle in the residual state was found to increase linearly with particle elongation.

A {\it three dimensional} rectangular column of particles was subjected to uniaxial compression in experiments and numerical simulations by Wiacek et al. \cite{Wiacek2012}. This study showed that an increase in particle aspect ratio strongly affects the mechanical response of the sample. The lateral-to-vertical pressure ratio decreased as the particle aspect ratio increased from unity to 1.6 and thereafter remained relatively unchanged with further increase in aspect ratio. In a recent numerical study by discrete element simulations platy particles were exposed to simple shear \cite{Boton2013}. These studies revealed an increase of the shear strength due to the alignment of the particles. Other numerical simulations using DEM investigated the behavior of monodisperse samples \cite{Ng2004} and bidisperse mixtures \cite{Ng2009,Ng2009b} under isotropic or triaxial compression. Friction angles obtained for bidisperse mixtures were found to be higher than for the mono-sized specimens, illustrating that increasing particle length lead to decreasing friction angles. For the bidisperse mixtures, the relationship between friction angle and void ratio was linear, and its slope changed with particle shapes. Other DEM
calculations \cite{Antony2004} of triaxial compression demonstrate that a single fabric measure can be defined, which shows strong correlation with the deviator stress for three different particle shapes (sphere, oblate and prolate ellipsoids).

Trepanier and Franklin studied the collapse of heaps of elongated grains \cite{Trepanier2010}. They varied
two different aspect ratios (particle length/diameter and pile height/radius). For particles with aspect ratio below 24, there exists a critical height $h_1$, below which the piles behave as a solid, they never collapse. Above a second critical height $h_2$, piles always collapse. Heaps with heights between these limits collapse with a probability which increases linearly with pile height. Long particles with aspect ratios exceeding 24
form piles that never collapse.
Several aspects of these experiments were confirmed by recent 2D DEM simulations by Tapia-McClung and Zenit \cite{TapiaMcClung2012}. Elongated grains were approximated by $k$ circular segments ($1\le k\le 8$) linearly glued together. After the collapse, both the final runout distance and height of the deposits scaled as a power law of the initial aspect ratio of the column, independent of the elongation $k$ of the particles. Collapsing heaps of elongated linear chains of grains behave qualitatively similar to those of circular grains.

\section{Fluidized state}

When energy is supplied to the system in the form of vibration, air flow, electromagnetic fields, etc.,
or when potential energy in the gravitational field is released during sliding or falling, granulates can be considered to be in a fluidized state. Traditionally, the term {\it fluidized granular material} is generally used to describe systems where most of the particles are permanently in contact with neighbors. At higher energy input, one can also achieve a more dilute a gas-like state, where particles interact only by independent binary collisions. This state will be addressed in Sec.~\ref{sec:gas}.

Technically, granular beds are often fluidized by a gas flow, where air is blown through the granulate to agitate, mix or dilute the granular bed. Mathematically, this represents a complex problem of two-phase flow.
A computational fluid dynamics DEM approach was applied by Zhou et al.~\cite{Zhou2011} to consider the gas fluidization of ellipsoidal particles and to identify differences with spherical grains. Particles were either oblate or prolate, aspect ratios varied from very flat (aspect ratio $>0.25$) to elongated (aspect ratio $<3.5$). The comprehensive simulations started with particles poured randomly into a bed.
The particle shapes were found to affect the bed permeability, which deteriorated much for oblate particles,
while beds of prolate particles were comparable to those of spheres.
The analysis of particle orientations showed that the bed structures for
ellipsoids are not random. Oblate particles prefer facing upward with the flat side, while
prolate particles prefer horizontal orientation of their long axis.

\subsection{Shaken grains}

\label{sec:shaken}

In many situations, granular material is fluidized by vertical shaking. Under these conditions, shape-anisotropic grains will generally develop an orientational order. This has been demonstrated in simulations as well as in experiments. Two particularly simple 2D systems are (i) horizontal or (ii) vertical monolayers of grains subjected to vertical shaking. Two phenomena distinguish systems of elongated grains from those made of spheres, the development of orientational order in the plane or vertical to it, and directed structures of motion in the plane.

\begin{figure}[htbp]
\centering
 \includegraphics[width=\columnwidth]{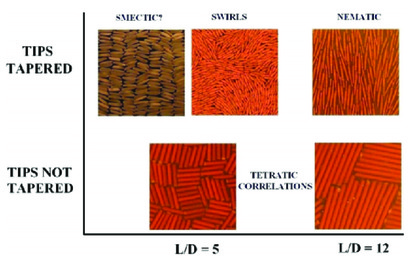}
 \caption{Schematic overview of steady states observed in a vibrated monolayer by Narayan et al.
 in the parameter space spanned by the aspect ratio and two different particle shapes.
Figure reprinted with permission from Ref.~\cite{Narayan2006}. Copyright (2006) by IOP.
}
\label{fig:narayan}
\end{figure}

A horizontal layer of densely packed elongated grains under vibration was investigated by Narayan et al. \cite{Narayan2006} to study orientational and positional order. The authors found a sensitive dependence of the ordered structures developed by the grains on details of the grain shape like aspect ratios and tapered or non-tapered tips. Monodisperse rods developed tetratic correlations, while sphero-cylinders showed an isotropic to nematic transition. For rice grains, smectic like order was observed (Fig.~\ref{fig:narayan}).

The isotropic-nematic transition was also captured in a kinetic theory developed by Ben Naim and Krapivsky \cite{BenNaim2006}. In the steady state, alignment by pairwise interactions is balanced by the diffusive motion. At large diffusivities, the system is disordered, while at low diffusivities it becomes ordered.

In a more dilute arrangement, an isotropic-nematic transition was observed in a vertically vibrated quasi-2D
confined horizontal layer of long rods ($L/D>20$) as a function of rod density \cite{Galanis2006,Galanis2010}.
In the nematic phase, increasing rod density led to a transition from bipolar ordering, dominated by the
bounding wall of the circular container, with two lateral defects in the orientational alignment structure near the walls, to uniform alignment.
A continuum free energy functional captures key patterning features
down to almost the particle size. The authors estimated relative bend
and splay elastic constants and wall anchoring strengths by fitting theory to experimental data.
For high particle densities, complex patterns emerged due to a competition between bulk and boundary alignment.

Large-scale collective motion was found in a monolayer of vertically vibrated elongated particles on a
horizontal dish \cite{Aranson2007}. Three types of rice with different aspect ratios,
metal dowels, and mustard seeds were compared. For rice grains,
recurring swirls with characteristic scales much larger than individual particle sizes were observed.
A small horizontal component of the vibration in combination with orientation-dependent bottom friction
was identified as the source of the swirl formation. An unexpected result was the very strong sensitivity of swirling motion to the particle shapes:
No swirling was found for monodisperse metal cylinders, very
little swirling occurred for long, thin (Basmati) rice grains. This could be a consequence of particle interlocking and the formation of tetratic structures with higher rigidity
and resistance to shear. Combining the velocity field and local alignment tensor, the authors developed a continuum model which qualitatively describes the experiment.

A similar experiment was reported by Blair et al.~\cite{Blair2003} who studied cylindrical rods with aspect ratio 12.4 and different filling fractions on a horizontal dish. The authors found a vortex motion at low filling fractions, where the rods were preferentially aligned in the horizontal orientation. For higher filling fractions, domains of vertically aligned rods occurred. The critical fill level was approximately half the particle density of a hexagonal arrangement consisting of vertically aligned rods.

In the same paper \cite{Blair2003}, the authors also reported experiments under horizontal shaking and with a single row of tilted rods in an annular container geometry to elucidate the role of tilt. The direction of motion in the plane was always in the tilt direction.
A similar geometry was investigated experimentally and by numerical modeling by Volfson et al.~\cite{Volfson2004}.
A set of nearly vertical rods confined to an annulus was shaken vertically.
The rods move horizontally, with their direction of motion depending on the driving acceleration.
Three types of motion were identified: slide, slip-stick, and slip reversal.
For large acceleration, the rods are transported in tilt direction, but for small acceleration they can move in the opposite direction.
Molecular dynamics simulations capture this behavior and reproduce the experimentally observed phase diagram.

The translational motion of elongated grains on a vibrating plane has been in the focus of several other experiments. Transport of elongated particles on a ratchet-shaped base was studied by Wambaugh et al. \cite{Wambaugh2002}. Monomers, dimers and trimers of collinear spheres were simulated on
a vibrating ratchet. Differences in layer velocity profiles and net grain velocities were observed for grains composed of one, two, or three spheres. In the case of mixtures with different species of grains, there is a layer-by-layer variation in the average velocity as well as a layer segregation of the species, and a horizontal separation of the species can be obtained.

An active nematic in 2D realized by shaking a container with a monolayer of apolar rods was reported by Narayan et al. \cite{Narayan2007}.
Giant number fluctuations (flocking) were found.
These fluctuations were first attributed to curvature-driven active currents specific for
nonequilibrium nematic systems. However, Aranson et al. \cite{Aranson2008}
 argue that their appearance could also be related to other observations made on
spherical particles, where giant number fluctuations arose either from dynamic inelastic clustering
or from a persistent density inhomogeneity.

For disks with built in polar asymmetry, large scale collective motion and
giant number fluctuations were observed experimentally in a vibrated monolayer \cite{Deseigne2010}.
Here, benchmark experiments have been carried out using the same setup with symmetric particles.
The latter show no collective motion,which confirmed that the observations with polar particles were
not caused by a residual horizontal component of the acceleration.
The scaling exponent associated with the giant number fluctuations was in accordance with earlier
theoretical predictions \cite{Ramaswamy2003,Chate2006}.

Self propelled polar rods show a tendency to aggregate near the walls since they are unable to
turn around and escape under low noise conditions. When the vibration strength (noise) is increased,
aggregation is reduced and a uniformly distributed state is observed, with local orientational
order and swirling \cite{Kudrolli2008}. Such aggregation was not observed for point-like or round
self-propelled
particles \cite{Chate2006,Deseigne2010}. The concentration dependent anisotropic diffusion was
also measured for self propelled polar rods \cite{Kudrolli2010}. The mean square displacement of the grains grows linearly with time for low filling fractions. Autocorrelations in the direction of
motion decay progressively slower with increasing filling ratios.

\begin{figure}[htbp]
\centering
 \includegraphics[width=\columnwidth]{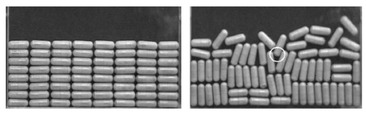}
 \caption{64 sphero-cylinders of 23 mm length and 8.1 mm diameter in a quasi-2D arrangement directly
 after filling (left) and after 1650 sec of vertical vibration with a peak acceleration of 3~$g$.
Figure reprinted with permission from Ref.~\cite{Ramaioli2007}. Copyright (2007) by the American Physical Society.
}
\label{fig:ramaioli}
\end{figure}

The second 2D geometry, a vertical quasi-twodimensional system of rods, has been studied experimentally as well as in DEM simulations by Ramaioli et al.~\cite{Ramaioli2005,Ramaioli2007}. The elongated rods showed a tendency to align vertically, as seen in Fig.~\ref{fig:ramaioli} (note the difference to Ref.~\cite{Zhou2011}). The alignment was found to occur regardless of initial conditions, and also independent of influences from side walls. An optimum acceleration to promote vertical ordering was found. The optimum excitation parameters depend on the aspect ratio of the particles. Short rods showed a lower tendency to align vertically than long ones. An interpretation was given in terms of an energy barrier for the
individual particles to leave a local lattice position.

A similar system was studied experimentally and theoretically by Lumay et al.~\cite{Lumay2006}. In a vertical monolayer of rods under vertical tapping, the compaction and formation of domains of aligned grains was observed. After nucleation, these domains experience some shearing and they develop an optimal order.
In a microscopic description, this complex process starts with translational motions of the grains,
while at later stages, rotations become more pronounced.

\begin{figure}[htbp]
\centering
 \includegraphics[width=6cm]{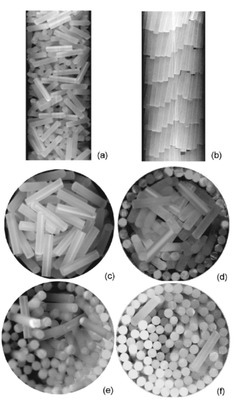}
 \caption{Snapshots of configurations of rods in a vertical shaken cylinder.
a) side view after filling,
b) relaxation after many thousands of taps,
c-f) evolution of the initial, disordered state to a highly
ordered, nematic state (material from the upper half column was removed).
d) after 2000 taps, e) after 6000 taps,
f) long term behavior.
Figure reprinted with permission from Ref.~\cite{Villarruel2000}. Copyright (2000) by the American Physical Society.
}
\label{fig:villarruel}
\end{figure}

A small-scale three-dimensional system of vertically shaken rods was introduced by Villarruel at al.
\cite{Villarruel2000} (Fig.~\ref{fig:villarruel}). Rods in a cylindrical container developed orientational (nematic) order when the system was shaken vertically. In this geometry, ordering was initiated at the container walls and propagated inwards.
When starting from an initially disordered configuration, the development of orientational order is associated with a steep rise in the local packing fraction.

Ribiere et al. \cite{Ribiere2005} performed a shaking experiment in a deep granular bed, Basmati rice (aspect ratio 2.5) was studied in a vertical cylinder of 10 cm diameter. It was filled to a height of 10 cm. The authors found two qualitatively different scenarios depending on the tapping intensity. At high acceleration, a single (vertical) vortex was observed, while at low tapping two unstable vortices appeared in the upper part of the container. Below them, an ordered compact zone was created, where the grains were oriented preferentially horizontally.

Experimental and numerical investigations of compaction in granular materials composed of rods was reported in a similar geometry by Lumay and Vandewalle \cite{Lumay2004}. As a function of the relative sizes of the particles and the container diameter, large variations of the asymptotic packing volume fraction are found, the relevant parameter being the ratio between the rod length and the tube diameter. Pasta noodles served as the granular material. The compaction dynamics were independent of the particle length. At a particle-length to tube-diameter ratio close to one, a transition between 3D and 2D ordering was observed. The main features of the experimental findings could be reproduced by numerical simulations using a simple toy model \cite{Lumay2004}.

The segregation of mixtures of grains with different shapes and sizes vertically shaken in a flat (Hele-Shaw) cell has been investigated experimentally and modeled in simulations by Caulkin et al.~\cite{Caulkin2010}. The authors studied crushed limestone grit, fine grain sand, rice, and pasta tubes in pseudo two-dimensional container geometry. Mixtures of non-spherical, arbitrary shaped/sized particles were prepared. After several thousand taps,  stratification in horizontal segregated layers with the different materials was observed. The experimental data were compared to results of computer simulations. The numerical data suggest that segregation can be explained by geometrical considerations, as a result of the relative motion between
particles of different sizes and shapes. A geometrical algorithm was suggested that provides a fast and qualitative prediction as to how likely segregation is to occur for a given mixture of grains.


\subsection{Flowing grains}

\label{sec:flowing}

Flow of grainy materials is often intrinsically connected to shear of the granular bed. Nevertheless, we
try to treat both processes in separate sections, the present one focused on transport of grains in flow processes, and
the following one dealing with shear induced phenomena like grain alignment and dilatancy.

The most frequently studied case for the liquid-like motion of anisometric grain assemblies is hopper flow.
This is probably due to the huge number of applications of this geometrical configuration in industry or agriculture.
Laboratory studies revealed that in hopper flows of elongated grains, the basic features are very different
from those of spherical particles \cite{Baxter1990,Cleary1999,Cleary2002,Sielamowicz2005,Jin2010}. For elongated grains,
the flow becomes increasingly concentrated in a narrow funnel above the hopper opening.

The effect of grain elongation and blockiness (grain angularity) has been studied in DEM calculations by
Cleary \cite{Cleary1999,Cleary2002} in {\it two dimensions}. On one hand, these simulations showed that increasing blockiness decreases the flow rate, but does not significantly alter the flow pattern itself, as compared to
the case of flowing circular particles. On the other hand, the authors found a considerable difference in the flow pattern
and flow rate between a polydisperse sample of ellipses with $L/D=5$ and beads (with the same polydispersity in size)
where the diameter of the beads was in the same range as the length $L$ of the ellipsoids.

Numerical simulations of {\it three-dimensional} model systems using spherical and corn-shaped particles also
show that the flowing region is narrower for corn-shaped particles \cite{Tao2010}. Thus, even minor deviations
from a spherical shape are essential for the flow characteristics. Experiments using a three-dimensional silo yielded
increasing discharge rates in the order of (i) hexahedron, (ii) sphere and (iii) ellipsoid grains \cite{Jin2010}.
DEM simulations adequately model these experiments. Two- and three-dimensional calculations by Langston
et al. and Li et al. \cite{Langston2004,Li2004} show larger flow rates for shape anisotropic particles (ellipses or disks)
compared to spheres \cite{Langston2004,Li2004}. However, an earlier 2D DEM study \cite{Cleary2002} is in sharp contrast
with the above findings, as it found larger flow rates for spheres than for elongated particles. All of the above numerical studies
used particles with identical volumes for the comparison of hopper flow rates. The different conclusions might be
attributed to differences in the system size or model details (e.g. friction).
Markauskas and Ka\v{c}ianauskas \cite{Markauskas2011} calculated hopper discharge rates of axisymmetric particles and
the angle of repose of the simultaneously formed pile using a DEM multi-sphere model. Simulations and experiments on rice were found to agree quite well when the rolling resistance was taken into account.
When the diameter of the orifice is not much larger than the grain size, silo discharge can be blocked by the
formation of arches, causing the granulate to jam. Zuriguel et al. \cite{Zuriguel2005} investigated this phenomenon for
spherical particles, rice, and lentils and showed that the shapes of the particles influence the probability of
arch formation and avalanche size distributions. However, proper scaling with an appropriately chosen equivalence
radius reveals that the qualitative characteristics are the same for all these materials.
The mean avalanche size $\langle s\rangle$ is described by
$$
\langle s\rangle \propto (R_{\rm c}-R)^{-\gamma}
$$
where an exponent $\gamma\approx 6.9$ was found to be nearly independent of particle shapes,
$R$ is the ratio of orifice and particle (square equivalent) diameters, $R_{\rm c}$ is a particle-dependent
critical radius ratio.


\begin{figure}[htbp]
\centering
 \includegraphics[width=5.5cm]{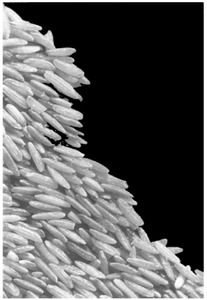}
 \caption{Close-up photograph of a rice pile by Frette et al. Most of the grains are preferentially aligned
 in flow direction, but with a considerable distribution of orientation angles.
 The local packing is inhomogeneous, and the slope is ragged.
 Figure reprinted with permission from Ref.~\cite{Frette1996}. Copyright (1996) by
 Nature Publishing Group.
 }
\label{fig:frette}
\end{figure}

Surface avalanches on granular heaps consisting of elongated grains (Fig.~\ref{fig:frette}) have attracted a lot of attention with the past two decades \cite{Frette1996,Gleiser2001,Ahlgren2002,Denisov2012}. The motivation for these studies was to check the validity of Self Organized Criticality (SOC), introduced by Bak, Tang and Wiesenfeld \cite{Bak1987}, for strongly nonspherical grains. Various materials have been studied including rice \cite{Frette1996,Ahlgren2002,Denisov2012}, lentils, quinoa, and mung beans \cite{Denisov2012}. These studies revealed that only the behavior of long grain rice piles complied the criteria for SOC, namely a power-law shaped avalanche size distribution, finite size scaling, and a universal scaling relation with certain characteristic exponents.

The influence of particle shape becomes particularly evident in the simple model system of connected dimers and trimers of beads studied by Olson et al. \cite{Olson2002}. One can associate specific types of granular flow with the particle geometries. In a 2D pile of simple spheres, which are stacked in a regular triangular lattice, particles flow regularly along lattice positions. In dimer piles, one observes a tumbling motion on the pile surface, while trimer piles are characterized by both tumbling at the pile top and disordered flow   throughout the bulk. The latter must obviously be attributed to the disordered, loosely packed arrangement of the trimers in the bulk.

\subsection{Sheared grains}

\label{sec:sheared}

A very simple model for the study of shear effects is a two-dimensional area filled with granular objects,
without gravitational forces, under periodic boundary conditions in Couette flow. Such a geometry was
considered by Cleary \cite{Cleary2008}, who simulated particles with various shapes (e.g. ellipsoidal or brick like). He found that the
aspect ratio and blockiness produce qualitatively comparable effects, but the aspect ratio has a larger influence. The shear strength of the material was found to increase strongly with
increasing aspect ratio. An increase by a factor of 25 to 31 in the collisional stress was observed between circular and 2:1 elliptical particles. The increased strength of the material has to be attributed to a better ability of the particles to interlock. This property determines whether parts will remain locked solid-like.
With increasing non-circularity of the particles, the sheared material becomes more homogeneous: the solid
fractions near the wall decrease with respect to circular particles, and in the middle of the cell it increases,
both fractions approach each other. In a certain range of aspect ratios, the velocity profile was found to become nearly
linear, and the wall slip almost disappeared.

El Shourbagy et al. \cite{ElShourbagy2006} considered a different geometry, they studied a two-dimensional
system under biaxial compression. The granulate is sandwiched between two plates that approach each other slowly
with constant velocity, while the lateral boundaries are assumed to provide constant pressure. In this geometry,
elongated particles also show a significantly higher shear strength than non-elongated particles. The Coulomb
friction coefficient plays a dominant role for the shape of the stress-strain characteristics. This indicates that the particle elongation must be considered as an important parameter in the statistical physics of granulates.

Computer simulations of the flow of elliptical particles with aspect ratios $L/D$ between 1 and 2 were reported by Campbell
\cite{Campbell2011}. The key result of the simulation is that ellipsoidal particles adopt a preferred
alignment in shear flow. Their major axis is inclined at an angle near $\arctan(D/L)$ with respect to
the main flow direction. The author points out that, perhaps coincidentally, this is the angle at which the
vertical extent of the ellipsoid is almost exactly equal to the the diameter of a sphere with the same volume.

Anki Reddy et al.\cite{AnkiReddy2009,AnkiReddy2010} considered a simple model of dumbbells in a 2D geometry.
They utilized numerical simulations to study the alignment of these particles in shear flow. The authors reported a continuous increase
in the alignment quality of sheared dumbbells with increasing aspect ratio and higher packing fraction,
but they found no discontinuity that might resemble a phase transition. The dumbbells were randomly oriented in the shear
flow at low packing fractions, and with increasing packing fraction, they developed a tendency to align towards the flow direction.
For dense flows of particles with aspects ratios between $1.25<L/D<2$, the average alignment angle $\theta_{\rm av}$
was in the range of $42^\circ>\theta_{\rm av}>30^\circ$.

\begin{figure}[htbp]
\centering
 \includegraphics[width=\columnwidth]{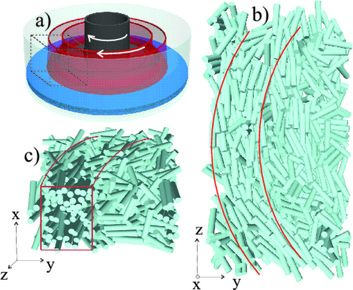}
 \caption{
a) Experimental geometry of the split-bottom cell.
A bottom disk having approximately 70~\% of the diameter of the cylindrical container is rotated
respective to the container and a ring-shaped shear zone forms above the disk edge.
The boundaries of the shear zone are indicated in the figure by red areas. The dashed box sketches
the volume evaluated in the X-ray CT.
b,c) Reconstruction of a typical arrangement of wooden pegs with 2.5 cm length and 0.5 cm diameter after
shearing in a split-bottom cell. Positions and orientations were calculated from
the X-ray CT data in the dashed region. Courtesy of S. Wegner  \cite{Wegner2012}.}
\label{fig:x-ray}
\end{figure}

The above numerical and experimental investigations on shear flows of shape-anisotropic granulates were restricted
to two-dimensions. There, a straightforward analysis of grain positions and orientations by means of optical imaging is possible.
{\it Three-dimensional} experiments require either the use of invasive observation methods (excavation) or tomographic tools.
Extensive quantitative experimental characterizations of the reorientation of elongated particles under shear
in a 3D geometry were recently performed by means of optical imaging and X-ray CT
\cite{Borzsonyi2012,Wegner2012,Borzsonyi2012b}. Grains with different aspect ratios and geometrical
shapes were sheared in a cylindrical split-bottom container (Fig.~\ref{fig:x-ray}a).
The particles in a free surface layer of the shear zone were detected optically, and their orientation was determined from video sequences,
from which statistics of the particle orientations were calculated. The same experiment was
performed with X-ray Computed Tomography, where the particles in the bulk of the granular bed could
be detected. The combination of both methods is very effective, since the expensive and time-consuming X-ray experiments
provide bulk data and allow the comparison of bulk and surface data to justify the relevance of surface data for statements
on the complete shear zone, and the optical imaging is a fast and inexpensive method to acquire extensive data sets of the particle alignment in the surface layer.

\begin{figure}[htbp]
\centering
 \includegraphics[width=\columnwidth]{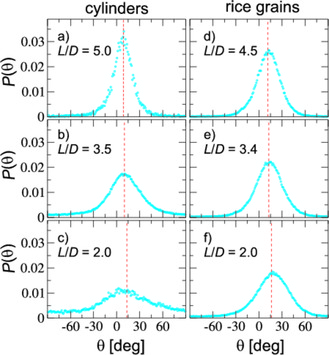}
 \caption{
 Distributions of the orientational angle
$\theta$ for cylindrical rods (left) and rice (right) with respect to the streamlines for
(a)-(c) cylinders with $L/D = 5.0$, 3.5, and 2.0, and (d)-(f) rice grains
with $L/D = 4.5$, 3.4, and 2.0, respectively. Data in panels (a)
and (c) were obtained by x-ray CT, other data were obtained by optical particle
detection on the surface. The vertical dashed lines mark the mean orientations, $\Theta_{\rm av}$.}
\label{fig:tamas3}
\end{figure}

Images ~\ref{fig:x-ray}b,c show the reconstruction of all grain positions in the observation
volume on the basis of the X-ray data (box sketched in Fig.~\ref{fig:x-ray}a).
The lines indicate the approximate boundaries of the shear zone. Inside the shear zone,
a preferential alignment of the grains has formed. This alignment can be evaluated quantitatively by computing the
local tensor {\bf T}:
\begin{equation}
T_{ij}= \frac{3}{2N} \sum\limits_{n=1}^N \left[{\ell}^{(n)}_i {\ell}^{(n)}_j -\frac{1}{3} \delta_{ij}
\right] \quad ,
\label{Q-tensor}
\end{equation}
where $\vec {\ell}^{(n)}$ is the unit vector along the long axis of particle $n$, and the sum is over
all $N$ detected particles.
The largest eigenvalue of {\bf T} is the primary order parameter $S$.
A second order parameter $S_{\rm b}$  is defined by the difference of the two other eigenvalues of {\bf T}. It describes biaxiality of the orientational order.
The directions of the principal axes of {\bf T} characterize the shear alignment.
Figure \ref{fig:tamas3} shows some typical results of measurements in the bulk (X-ray CT) and at the surface of the shear zone
(X-ray CT and optical imaging). As one can see, the long axes clearly exhibit a preferential alignment
to the local flow lines. Here, $\theta$ is the angle between the tangential flow direction and the local projection
of the long axes on the (practically horizontal) plane of flow and flow gradient. The distributions are narrower for
large aspect ratio particles, and wider for lower aspect ratio grains. Particularly for long grains, the
distributions can be satisfactorily approximated by Gaussians. The distributions are not centered about the flow
direction but the mean angle $\Theta_{\rm av}$ is of the order of $7^\circ$ to $17^\circ$
for aspect ratios between 2 and 5. The differences between cylindrical pegs and more ellipsoidal
rice grains at a given aspect ratio are not significant.
%

\begin{figure}[htbp]
\centering
 \includegraphics[width=\columnwidth]{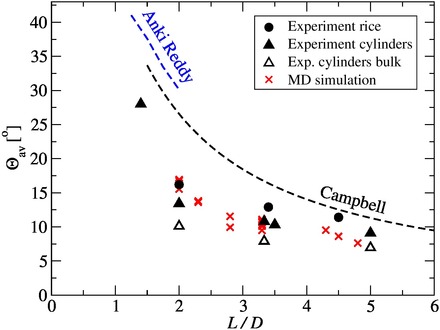}
 \caption{The average orientational angle $\Theta_{\rm av}$ as
a function of the particle aspect ratios $L/D$
obtained in experiments (bullets, triangles) and numerical simulations
(crosses). The dashed lines depict the relation given by Campbell \cite{Campbell2011} and
numerical data obtained for dense shear flows by Anki Reddy et al. \cite{AnkiReddy2009}}.
\label{fig:tamas6}
\end{figure}

The orientational order as well as the average alignment angle have been shown to reach stationary values
independent of the shear rate over three orders of magnitude of the latter.
Order parameters $S$ are found to be in the range between $\approx 0.4$
and $\approx 0.8$ both in the bulk and at the surface.
The biaxial order parameter $S_{\rm b}$ is much smaller, but is clearly nonzero.
Figure \ref{fig:tamas6} shows the mean alignment angles for grains of different shapes and
aspect ratios. The curve suggested by Campbell for elliptical particles in 2D \cite{Campbell2011} and the simulation data
of Anki Reddy et al. \cite{AnkiReddy2009} are added for comparison.
It was demonstrated in Ref.~\cite{Borzsonyi2012} that the relations between orientational order of the
rods and alignment angle are very similar to the values found in nematic liquid crystals
(see e. g.~\cite{Skarp1979,Gaehwiller1972,Beens1985,Ehrentraut1995}).
Numerical simulations with DEM \cite{Borzsonyi2012,Borzsonyi2012b} are in good agreement with the experiments.

Torque measurements of sheared elongated grains showed that
the alignment of the particles reduces friction in the sheared zone.
Starting from an initially random configuration of grains, the torque
drops by about 15 to 30~\% during the formation of alignment in the shear zone. The drop is stronger for higher aspect ratio particles \cite{Borzsonyi2012,Borzsonyi2012b}.
This is in accordance with recent DEM simulations by Guo et al. \cite{Guo2012}, where elongated particles
were modeled by cylinders or glued spheres (resulting in bumpy surfaces). These simulations covered a wide
range of densities, and aspect ratios from one to six. In the dense regime, where collision stresses dominate,
stresses were found to increase with decreasing particle aspect ratio and increasing bumpiness.

Further experiments were performed to observe the individual particle dynamics. It was established that
the elongated particles under shear perform in average a continuous rotation, with angular velocities
strongly dependent on the orientation relative to the flow field (in accordance with the stationary
distribution functions in Fig.~\ref{fig:tamas3}) \cite{Borzsonyi2012,Borzsonyi2012b}.
When the direction of the shear flow is reversed, the average alignment angle also reverses its sign.
The new preferential alignment of the particles is established when the shear strain $\gamma$ reaches approximately 5, i.e. when each particle in the shear zone has passed a couple of neighbors.

Another aspect of shear induced effects is the motion of objects through a densely packed assembly of grains,
the induced local realignment of grains and the involved drag forces.
The response of a three dimensional packing for large aspect ratio prolate particles against
local perturbation has been tested by Desmond and Franklin \cite{Desmond2006}.
They measured the force exterted on a small ball that was pulled vertically upwards.
For smaller aspect ratios and in large containers, granulate-like
behavior was observed, where the motion of the small ball induced local rearrangements in
the packing. For grains with large aspect ratio and smaller container size, solid-like behavior
was found, where the pulled ball lifted the whole granular column above it, like a solid body.
The power spectra of the force fluctuations, originating from the local rearrangements or stick
slip with the walls, showed power-law behavior with different exponents in the two cases.

Shear alignment of shape-anisotropic granular materials is a
long-known phenomenon in geophysical sciences. It becomes relevant in transport processes
of clasts that are carried or supported in an interstitial fluid, e.g. water in dense debris flows
or hot air in pyroclastic flows \cite{Karatson02}. Imbrication, viz the toppled-domino like arrangement
of anisotropic sediments, is often employed to determine the flow directions during such deposition processes.
This technique was described by Elston and Smith \cite{Elston1970} and applied in many subsequent
geological papers (e.g.~\cite{Smith1972,Rhodes1972}).
Even though such pyroclastic or debris flows may be turbulent (eddies may occur), a parallel-to-flow orientation
of the longest extension of clasts often results from the significant shear stress between
particles. Traditionally, such measurements were performed on the basis of manually selected clasts in a
vertical cross section of the sediment layers.
Kar\'atson et al.~\cite{Karatson02} developed a photo-statistical method
to quantify the strength of directional clast fabric, i.e. the vector length of clast alignment on
vertical outcrop faces, to determine orientational distributions
of various types of volcaniclastic mass-flow deposits.
Clast alignment was found in these investigations to be generally independent of clast sizes.

Porphyroclasts, i.~e. inclusions of clasts or mineral fragments in a metamorphic rock consisting of finer grained
crystals, are often rotated by shear stress if the embedding rock is strongly deformed.
Their shapes and orientations can be used as shear direction indicators.
Passchier \cite{Passchier1987} described the rotation of porphyroclasts under shear flow in dependence upon their elongation.
He distinguished two cases: porphyroclasts with aspect ratios close to one which permanently rotate in the sheared matrix,
and particles with large aspect ratios which reach a stable orientation.
Their role as shear sense indicators was discussed in detail in Ref.~\cite{Passchier2006}.
Mineral fish, i.~e. mineral grains or clusters of grains with typically asymmetric
sigmoid, lenticular and parallelogram shapes, are other inclusions often found in sheared rocks.
They can also be useful as shear sense indicators in geology \cite{Mukherjee2011}, but in addition to
shear stress, several other processes, such as dynamic recrystallization, must be taken into account, too.
The behavior of porphyroclasts under viscous shear flow was modeled in experiments with mantled solid spheres and
viscous spheres in PDMS \cite{Passchier1993}.

Ten Grotenhuis et al. modeled the shear alignment of mica fish and tourmaline fish by means of simplified laboratory experiments.
They studied anisometric objects in sheared matrices, such as rigid elongated objects with rhomboid cross sections embedded in Tapioca pearls \cite{Grotenhuis2002}.
With increasing aspect ratio, the long sides of the rhomboids were found to align to the shear bands in
the embedding granulate. Experiments with similar objects in a viscous fluid (PDMS), performed for comparison,
yielded completely different results. In PDMS, the aspect ratio was the dominant factor that determined the rotation of
the objects under shear, whereas in the granular bed, the elongated objects reached a stable orientation under shear, with
alignment angles below $25^\circ$ towards the flow direction.
Fish with small aspect ratio exhibit a larger angle than fish with large aspect ratio \cite{Grotenhuis2002,Grotenhuis2003}.

Another geological and technological field where the anisotropy of the individual grains is relevant
is soil mechanics.
Here, inherent and induced anisotropy are distinguished: the former is the result of the
sedimentation of particles, the latter occurs as a product of inelastic deformations.
Shape anisotropy of grains plays a crucial role for the fabrics, stress anisotropy and structure of voids, e.g.
in sands.
By means of optical microscope images, Oda et al.~\cite{Oda1982,Oda1985}
studied the orientations and positional relations of sand grains in thin sections of sand.
Three sources contribute to inherent anisotropy, viz. an anisotropic distribution of particle
contacts, the shape and orientational distribution of voids, and the
preferred orientation of non-spherical grains \cite{Oda1985}.
The principle axes of the fabric tensor and the stress tensor tend to align.
The emerging fabric is determined by both the particles' shape and by the deposition or compaction method.
Biaxial compression has been studied by discrete element simulations for elliptic cross-section rods
\cite{Shodja2003}. For random, mixed-size assemblies, the number of rolling contacts is larger than that of
sliding contacts. Rolling becomes more pronounced with increasing inter-granular friction angles.
These findings agree with Oda's experimental results \cite{Oda1982}.

The influence of particle shape
on the overall plastic response in sheared dense granular media was studied by Pe\~na et al.~\cite{Pena2007}. They employed 2D MD simulations in shear cells with periodic boundary conditions.
Ensembles of isotropic and elongated polygons were compared.
The dependence of the mechanical behavior on the evolution of inherent anisotropy was studied,
and differences between isotropic and elongated particles were discussed.
At large shear deformations, samples of grains with a given shape reach stationary values for both shear
force and void ratio independent of their initial configurations.

\section{Gaseous state}
\label{sec:gas}

When the density of particles per volume is very low, so that most of the time the grains move freely and particle contacts are reduced to rare collisions, then the granular system can be considered as a gaseous state. Since collisions in real granular systems are dissipative, a permanent kinetic energy supply is needed to maintain a stationary particle dynamics. A gas-like state of shape-anisotropic grains has been achieved in experiments e. g. by vibro-fluidization \cite{Atwell2005,Harth2011,Harth2013} or in an upflow of air \cite{Daniels2009,Daniels2011}.
As we will see, shape anisotropy of the grains matters in many respects.

Before we consider real granular gases, we introduce some experiments that deal with individual particle dynamics. These experiments give some insight in the energy and momentum transfers within loose granular ensembles, in particular when rotational energies become relevant.

\subsection{Single particles in a bath}
\label{sec:single}

An interesting issue with fluidized grains is the equipartition of energy among the degrees of freedom. The simplest experiment to characterize the dynamics of shape-anisotropic particles in a granular system is a single needle-shaped tracer particle immersed in a (2D) bath of point-like particles. In such a system, one can derive translational and rotational granular temperatures as a function of the mass ratio of the points and the needle, and the moment of inertia of the needle \cite{Viot2004}, by analyzing the needle motion statistically. These temperatures are, in general, not equal and they differ from the bath temperature. Only for very light bath particles is equipartition obtained asymptotically, regardless of the normal restitution coefficient.
A slight complication of this experiment is the steady-state dynamics of a granular capped rectangle (the 2D analogue of a sphero-cylinder) placed
in a 2D bath of thermalized hard disks \cite{Gomart2005}. Hard core collisions were assumed elastic between disks and inelastic between the capped rectangle and the disks.
With a Gaussian ansatz for the probability distribution functions, analytical expressions for the
granular temperatures were derived and the absence of equipartition was demonstrated.
The anisotropy and length of the capped rectangle, as well as the size ratio of the bath particles, were discussed and found to be important factors in determining whether equipartition can be expected.
Specifically, equipartition can be achieved in this system only with a particular non-trivial
choice of different restitution coefficients for the linear sides and the cap.

A simple planar granular rotator with a fixed center subject to inelastic collisions with bath particles was analyzed both numerically and analytically by Piasecki et al. \cite{Piasecki2006,Piasecki2007}. The angular velocity distribution evolves from quasi-Gaussian in the Brownian limit of a heavy rotator to an algebraic decay, when the rotator is much lighter than the bath particles \cite{Piasecki2007}.

\subsection{Experiments with granular gases}
\label{sec:expgaga}

As in the solid and the dense fluidized states, the realizations of granular gases can be classified into 2D and 3D geometries. Under normal gravity, one can realize 2D granular gases in an upward air flow on a horizontal plane. Mechanical excitation is possible, e.g., by shaking side walls of a lateral frame. Such a two-dimensional gas consisting of granular rods was studied experimentally by Daniels et al. \cite{Daniels2009}. Cylindrical plastic dowel pins with an aspect ratio close to four were fluidized in upflowing air. The area fraction chosen was 42\% so that a quasi-two-dimensional monolayer of apolar granular rods was formed. Under these conditions, the particles were uniformly distributed across the plane, without long-range ordering. Of course, rotational and translational motions couple by the collisions. The experiments yielded two results that are especially related to the particle shape: dynamical anisotropy, and superdiffusive dynamics in the direction
of the rod's long axes. The latter might be related to the special driving mechanisms in these experiments, because the upstream of air simultaneously serves as a lubricant to reduce friction and it supplies energy to maintain the dynamic equilibrium of the gas. The time scales where the mean square displacements parallel and perpendicular to the rod axes became diffusive differ. In a subsequent study, the authors reported propagating compression waves \cite{Daniels2011}. The speed of the collective waves was found to be an order of magnitude faster than the average particle. In these waves, large local particle number densities can occur.

Another driving mechanism was employed by Atwell and Olafsen \cite{Atwell2005}. They excited dumbbell-shaped
particles consisting of two loosely coupled spherical beads by means of a vertical vibration. The 2D dynamics, velocity fluctuations, and particle-plate interactions were investigated \cite{Atwell2005}, and an anisotropic behavior of the dimers was reported. The dimer motion is not strictly two-dimensional and the interactions are complex. Depending on the shaking parameters, the shape of the velocity distribution could be ''tuned'' from nearly Gaussian to nearly exponential.

\begin{figure}[htbp]
\centering
 \includegraphics[width=\columnwidth]{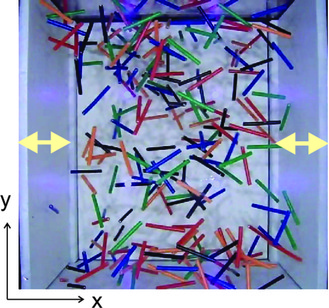}
 \caption{Snapshot of a granular gas of 250 cylindrical particles with lengths of 1.2 cm and diameters of 0.13 cm (insulated copper wires) in a box with three shaken side walls left, right (arrows) and rear. Photo courtesy of K. Harth.}
\label{fig:gagabox}
\end{figure}

Recently, experiments with a granular gas of elongated cylindrical particles were performed in a suborbital flight and in drop tower experiments \cite{Harth2011,Harth2013}. In these experiments, a truly 3D granular gas was realized and the dynamics of rotational and translational motions was extracted from video sequences. The gas was excited by shaking lateral side walls. Figure \ref{fig:gagabox} shows a snaoshot of the excited gas. Since the mean free path length in these experiments was not negligible compared to the container dimensions, the excitation is reflected in the velocity distributions. The distributions of the velocity component in the ''excitation'' direction $x$, with the shaken walls, contains pronounced high-velocity tails. ''Hot'' particles that are not yet equilibrated after momentum input from the side walls generate these tails. On the other hand, the distribution of the velocity component $v_y$ is roughly Gaussian, but more accurately described by a distribution $P(v_y) \propto \exp (-|v_y|^{1.5}/C^{1.5})$.
The rotational velocities of the particles were also evaluated and a large difference between the energies contained in the translational motion along $y$ and the rotational degrees of freedom was reported \cite{Harth2013}.

\subsection{Modeling of granular gases}
\label{sec:simgaga}

Since the experimental implementation of granular gases is often not straightforward, particularly in 3D,
a majority of studies on anisotropic particles in dilute ensembles employ numerical simulations.
Granular gases are somewhat easier to treat numerically than condensed granular phases because of lower
particle number densities and the nature of the interactions.
Computation time can be optimized with event driven computation schemes. These algorithms take advantage
of the fact that particle interactions are restricted to short collisions. After a collision,
velocities are updated with a certain coefficient of restitution. In intervals between collision events,
the particles are allowed to move freely.

A one-dimensional dilute gas of inelastic rods was modeled with a hybrid Monte Carlo and event driven
algorithm \cite{Aspelmeier1998}. Velocities were updated after collisions with a random coefficient of
restitution. A stochastic energy transfer towards internal vibrations of the grains was allowed, and
the latter are assumed to decay on a time scale that is slow compared to the average time between particle
collisions. Under these assumptions, an inelastic collapse was avoided. The complex cluster dynamics during
cooling was simulated. The long term dynamics and evolution of clusters were studied.
Large clusters of particles were observed to form, collide, merge or dissolve. Inside the clusters, an
equilibrated Maxwell velocity distribution of particle velocities was found, with nonzero mean velocities.
Among different clusters, the parameters of the distributions varied.

Chrzanowska and Ehrentraut \cite{Chrzanowska2002,Chrzanowska2003-1} studied hard needles in 2D.
They reported an occurrence of orientational clusters (bundles) for a cooling gas in molecular dynamics
simulations. A nematic phase was observed above a certain particle density $\rho^\ast$, which was
defined as the ratio of the area for a square with sides of the particle length $L$ and the area available
per particle. The simulations were reported in the nematic state at $\rho^\ast=8$. The clusters emerged
in the isotropic as well as in the nematic states.

Further simulations of a 2D gas of hard needles by event-oriented MD \cite{Foulaadvand2011} provided translational
and rotational diffusion constants and their dependence on density. In literature, one finds a continuous spectrum
of papers simulating interacting hardcore particles in a wide range of densities, which describe many different particulate systems, from dilute granular gases to liquid crystals \cite{Frenkel1987,Kroger2004,BenNaim2006}.
For example, the paper by Anki Reddy et al. \cite{AnkiReddy2009} mentioned above in the context of shear effects
of fluidized granulates, describes shear alignment as a function of packing densities
in a broad range from gas-like to liquid-like granular behavior.
In the following section, we restrict to the compilation
of effects in very dilute systems far from the condensed phase.

A simple geometrical model for gases of anisometric particles is that of inelastic dimers.
A vertical, two-dimensional vibrated bed of dumbbells was investigated using
experiments and numerical simulation \cite{Wildman2009}. Experimentally,
the location of the center of mass, and the locations and direction of motion of
the component particles of the dumbbells, were determined. Numerically, equations of motion for each of
the particles were solved and compared to the experiment. The authors suggested that Gaussians represent
reasonable approximations of the velocity distributions, except close to the boundaries.
A similar 2D system has been considered by Costantini et al. \cite{Costantini2005}.
Their simulations confirmed Haff's law for the decay of the total energy of granular gases with $t^{-2}$.
The simulations with different restitution coefficients yielded Gaussians for the distributions of rotational
and translational energies. The equipartition was satisfied only in the case of fully elastic collisions.

Another 2D simulation of the cooling dynamics of a granular gas of elongated particles was performed by Kanzaki et al.~\cite{Kanzaki2010}.
The particles used in the simulations were irregular convex polygons generated by a
random Voronoi tessellation. They authors reported two different scenarios: For weakly dissipative collisions,
a homogeneous cooling process was observed. The overall translational kinetic energy decreases in the same way
as for viscoelastic circular particles. For strongly dissipative collisions, cooling is inhomogeneous.
The rotational kinetic energy decays in agreement with Haff's law, while the decay of the translational
energy slows down. Equipartition is violated during the cooling process. It is assumed that strong dissipation
and particle anisotropy lead to the formation of ordered cluster structures which retard the relaxation.

Huthmann et al. \cite{Huthmann1999} used kinetic theory and event driven dynamics simulations to examine the free
cooling of granular needles in three dimensions. They found an exponentially fast initial decay towards a state
where the ratio of translational and rotational energies is constant, followed by an algebraically slow decay where the
constant ratio is maintained. The ratio of translational to rotational energies reported in the study depends
on the coefficient of normal restitution and on the mass distribution along the needles.

Villemot and Talbot \cite{Villemot2012} extended this subject to the
homogeneous cooling of hard, smooth ellipsoids in three dimensions. An event driven
numerical simulation was performed. The elongation of the particles was found to have a
strong effect on the cooling characteristics. Weakly elongated ellipsoids
displayed two distinct cooling regimes. At short times, translational and rotational energy
were found to decay at different rates until their ratio reached a time-independent value.
Equipartition was violated. The overall temperature decays as predicted by
Haff's law, $\propto t^{-2}$. For more elongated ellipsoids, translational and rotational
temperatures rapidly reached a constant ratio near unity. Velocity distributions were found
to be Gaussian with insignificant deviations towards larger values at high velocities.

\section{Summary and outlook}

Granular physics as a broad and complex field of soft matter physics still provides challenges for further research.
Even the behavior of spherical or nearly shape-isotropic grains is still not completely resolved.
Particles of anisotropic shape, even in their simplest form of prolate or oblate axially symmetric grains,
add much more complexity to the static and dynamic characteristics of the ensembles,
and lead to qualitatively new macroscopic phenomena.
Grain shapes often have a much larger influence on packing and dynamics than details
of the particle surfaces and interactions.
In most situations, the statistics of a large number of particles in the ensemble do not lead to the averaging of the constituents' anisotropic character
Instead, the very preparation of the granular bed, interactions of grains with container walls, or even a slight agitation of the ensemble by shaking or shearing induces particle alignment. Such an alignment, in turn, can alter the macroscopic properties of the ensemble.

To date, there is still a lack of quantitative experiments, in particular with (3D) bulk samples. The main reason for that is the difficulty in characterizing particle configurations inside the granular bed. This task
involves either destructive methods or requires the application of expensive X-ray or Nuclear magnetic resonance tomographic techniques. Thus, many studies have been restricted to 2D geometries. Simulations are often much more complicated than with spherical grains. Often, grain shapes are approximated in 3D numerical studies by chains of spheres, ellipsoids or (sphero-) cylinders, and in 2D by chains of circles, ellipses or rectangles.
It has been demonstrated that in some situations secondary shape details other than aspect ratios can have dramatic influences on the structure of granular packings (e. g. \cite{Schreck2010,Shen2012}). Therefore, one may expect further surprising details from simulations of particles with more complex geometries.

Previous investigations already revealed a number of well-established properties and phenomena of shape-anisotropic granulates: In static assemblies, the most important findings are differences in the packing and jamming behavior compared to spherical particles. Random packing of ellipsoids leads to higher packing densities than for spheres \cite{Donev2004}, and two maxima in the packing fraction versus aspect ratio characteristics are found, one for oblate ellipsoids, and one for moderately prolate ones. For very long cylinders, the random close packing density goes to zero, as it is inversely proportional to the aspect ratio.
Mixtures of spheres and short sphero-cylinders show a behavior unusual for granulates: they behave isochoric when they are mixed \cite{Kyrylyuk2011b}.

In general, orientational order leads to a compaction of anisotropic grains. Therefore, the determination of random packing is not trivial. Filling containers in a gravitational field by pouring particles can already lead to partial alignment and higher packing fractions \cite{Buchalter1992}. Shaking the container can furthermore increase alignment and compaction, both for prolate and oblate particles.

Force networks and distributions of stress in heaps or containers depend upon the aspect ratio and state of alignment of grains. Their characteristics can significantly influence the statics of granular piles. For example, the stability of piles is crucially dependent upon the aspect ratio of long objects.
Ensembles of very long rods exhibit a solid-like shape stability \cite{Philipse1996,Trepanier2010}.

Particle shapes do not only quantitatively change the dynamics of flowing grains; they also add new peculiarities. Avalanches observed in piles of long-grained rice reproduced the criteria for self-organized criticality \cite{Frette1996}, a phenomenon long sought for in avalanches of isometric granulates. Under shear flow, orientational order is induced in the shear zone of anisometric grains \cite{Borzsonyi2012}, both for elongated and flattened particles. The average orientation of the particles forms a small angle with the flow direction, a phenomenon that appears to be very similar, even quantitatively, to molecular liquid crystals. The shear alignment angle decreases with particle aspect ratio. Partial alignment in the shear zone has crucial effects on the shear stress. The shear stress of an initially unaligned sample decreases by up to 30~\% in the course of the establishment of alignment.

An interesting aspect is the dynamics of granular gases of elongated, rod-like particles, which represent in some respect an analogy to a two-atomic gas: the individual particles have three translational and two rotational degrees of freedom, and the kinetic energy is distributed among them. Such ensembles represent unique mechanical systems to study statistics of large ensembles composed of interacting particles.
Both numerical simulations \cite{Kanzaki2010,Villemot2012} and experimental results \cite{Harth2013} suggest that the equipartition of kinetic energies on the individual degrees of freedom is violated in general. More experimental quantitative results are needed to establish velocity distribution functions with high accuracy in order to compare experiments and theoretical results reliably and to test the predictions of existing models.
Valuable insight may be obtained from experiments with granular gases during cooling, in absence of excitations.

This review concentrates only on certain aspects of grain shape anisotropies, we have mainly focused on systems with elongated or flattened grains. Many other parameters of non-sphericity can become relevant for packing and flow properties, too. Such aspects have been explored, for example, by the CEGEO group \cite{CEGEO} in 2D systems.

In 3D, most of the experiments and simulations dealt with axially symmetric, apolar grain shapes so far. It is interesting to compare this situation with molecular liquid crystal phases. There, systematic research started in the beginning of the 20th century with the nematic phase of calamitic (rod-shaped) mesogens, which is axially symmetric and apolar. Meanwhile, numerous discotic (disk-shaped) mesogens have been synthesized and extensively studied. But also sanidic (board-shaped) mesogen structures have attracted interest, in first line in the search for biaxial nematics. Chiral phases have become attractive primarily because of the polar (ferroelectric) properties of
chiral smectics \cite{Meyer1975}. Since nearly 20 years, the discovery of bent-core mesogens boosted the research on phases which break chiral symmetry spontaneously \cite{Takezoe2006}. Dozens of new liquid crystalline phases have been described, and the study of their dynamic and structural properties is a flourishing area of research.

Granular materials, on the other hand, are only at the beginning of such a tremendous research activity. There is a rich potential for future investigations. Board-like shapes may combine features of prolate and oblate grains. They introduce more complexity than rods or axially symmetric ellipsoids, for example one can expect a large biaxiality in the orientational order parameter. Chirality can be introduced with screw-like objects, cone-shaped or bent grains can introduce polarity in packing structures, in orientational order or in dynamic properties.

A challenging task could be the search for granular systems that develop smectic-like layer ordering in 3D bulk samples, such as observed in 2D or small-scale 3D systems (Figs.~\ref{fig:narayan},\ref{fig:ramaioli}, and \ref{fig:villarruel}). It is well known that in nematics with pre-smectic fluctuations, the shear alignment is completely different from the 'standard' behavior with an asymptotic alignment angle. In such systems, a tumbling regime is observed instead \cite{Carlsson1986}.

The topic reviewed here is not exclusively of academic interest. There exist many connections between granular physics in the laboratory and various related fields like geophysics, agriculture or transport technology, where shape-anisotropic grains are encountered. In these fields, much empirical knowledge has been collected and is exploited in practical tasks. A quantitative understanding of the relationship between particle shapes and their behavior in macroscopic ensembles may thus have a substantial impact on technology and our control of natural processes.

\section*{Acknowledgments}

The authors acknowledge financial support by the DAAD/M\"OB researcher exchange program (Grant No. 29480).
Kirsten Harth is acknowledged for discussions, literature references, and providing a figure.
Sandra Wegner is acknowledged for contributing a figure.
We thank Tanya Ostapenko for proofreading.

\providecommand*{\mcitethebibliography}{\thebibliography}
\csname @ifundefined\endcsname{endmcitethebibliography}
{\let\endmcitethebibliography\endthebibliography}{}


\begin{mcitethebibliography}{170}
\providecommand*{\natexlab}[1]{#1}
\providecommand*{\mciteSetBstSublistMode}[1]{}
\providecommand*{\mciteSetBstMaxWidthForm}[2]{}
\providecommand*{\mciteBstWouldAddEndPuncttrue}
  {\def\EndOfBibitem{\unskip.}}
\providecommand*{\mciteBstWouldAddEndPunctfalse}
  {\let\EndOfBibitem\relax}
\providecommand*{\mciteSetBstMidEndSepPunct}[3]{}
\providecommand*{\mciteSetBstSublistLabelBeginEnd}[3]{}
\providecommand*{\EndOfBibitem}{}
\mciteSetBstSublistMode{f}
\mciteSetBstMaxWidthForm{subitem}
{(\emph{\alph{mcitesubitemcount}})}
\mciteSetBstSublistLabelBeginEnd{\mcitemaxwidthsubitemform\space}
{\relax}{\relax}

\bibitem[Coulomb(1773)]{Coulomb1773}
C.~A. Coulomb, \emph{Acad. R. Mem. Phys. Divers Savants}, 1773, \textbf{7},
  343\relax
\mciteBstWouldAddEndPuncttrue
\mciteSetBstMidEndSepPunct{\mcitedefaultmidpunct}
{\mcitedefaultendpunct}{\mcitedefaultseppunct}\relax
\EndOfBibitem
\bibitem[Liu \emph{et~al.}(1995)Liu, Nagel, Schecter, Coppersmith, Majumdar,
  Narayan, and Witten]{Liu1995}
C.-H. Liu, S.~R. Nagel, D.~A. Schecter, S.~N. Coppersmith, S.~Majumdar,
  O.~Narayan and T.~A. Witten, \emph{Science}, 1995, \textbf{269}, 513\relax
\mciteBstWouldAddEndPuncttrue
\mciteSetBstMidEndSepPunct{\mcitedefaultmidpunct}
{\mcitedefaultendpunct}{\mcitedefaultseppunct}\relax
\EndOfBibitem
\bibitem[Bak \emph{et~al.}(1987)Bak, Tang, and Wiesenfeld]{Bak1987}
P.~Bak, C.~Tang and K.~Wiesenfeld, \emph{Phys Rev. Lett.}, 1987, \textbf{59},
  381\relax
\mciteBstWouldAddEndPuncttrue
\mciteSetBstMidEndSepPunct{\mcitedefaultmidpunct}
{\mcitedefaultendpunct}{\mcitedefaultseppunct}\relax
\EndOfBibitem
\bibitem[Bak \emph{et~al.}(1988)Bak, Tang, and Wiesenfeld]{Bak1988}
P.~Bak, C.~Tang and K.~Wiesenfeld, \emph{Phys Rev. A}, 1988, \textbf{38},
  364\relax
\mciteBstWouldAddEndPuncttrue
\mciteSetBstMidEndSepPunct{\mcitedefaultmidpunct}
{\mcitedefaultendpunct}{\mcitedefaultseppunct}\relax
\EndOfBibitem
\bibitem[Reynolds(1885)]{Reynolds1885}
O.~Reynolds, \emph{Philos Mag. Ser. 5}, 1885, \textbf{20}, 469\relax
\mciteBstWouldAddEndPuncttrue
\mciteSetBstMidEndSepPunct{\mcitedefaultmidpunct}
{\mcitedefaultendpunct}{\mcitedefaultseppunct}\relax
\EndOfBibitem
\bibitem[Helbing(2001)]{Helbing2001}
D.~Helbing, \emph{Rev. Mod. Phys.}, 2001, \textbf{73}, 1067\relax
\mciteBstWouldAddEndPuncttrue
\mciteSetBstMidEndSepPunct{\mcitedefaultmidpunct}
{\mcitedefaultendpunct}{\mcitedefaultseppunct}\relax
\EndOfBibitem
\bibitem[Chowdhury \emph{et~al.}(2000)Chowdhury, Santen, and
  Schadschneider]{Chowdhury2000}
D.~Chowdhury, L.~Santen and A.~Schadschneider, \emph{Phys. Reports}, 2000,
  \textbf{329}, 199\relax
\mciteBstWouldAddEndPuncttrue
\mciteSetBstMidEndSepPunct{\mcitedefaultmidpunct}
{\mcitedefaultendpunct}{\mcitedefaultseppunct}\relax
\EndOfBibitem
\bibitem[Nagatani(2002)]{Nagatani2002}
T.~Nagatani, \emph{Rep. Prog. Phys.}, 2002, \textbf{65}, 1331\relax
\mciteBstWouldAddEndPuncttrue
\mciteSetBstMidEndSepPunct{\mcitedefaultmidpunct}
{\mcitedefaultendpunct}{\mcitedefaultseppunct}\relax
\EndOfBibitem
\bibitem[Helbing \emph{et~al.}(2005)Helbing, Buzna, Johansson, and
  Werner]{Helbing2005}
D.~Helbing, L.~Buzna, A.~Johansson and T.~Werner, \emph{Transportation Sci.},
  2005, \textbf{39}, 1\relax
\mciteBstWouldAddEndPuncttrue
\mciteSetBstMidEndSepPunct{\mcitedefaultmidpunct}
{\mcitedefaultendpunct}{\mcitedefaultseppunct}\relax
\EndOfBibitem
\bibitem[Biham \emph{et~al.}(1992)Biham, Middleton, and Levine]{Biham1992}
O.~Biham, A.~A. Middleton and D.~Levine, \emph{Phys. Rev. A}, 1992,
  \textbf{46}, R6124\relax
\mciteBstWouldAddEndPuncttrue
\mciteSetBstMidEndSepPunct{\mcitedefaultmidpunct}
{\mcitedefaultendpunct}{\mcitedefaultseppunct}\relax
\EndOfBibitem
\bibitem[Nagel and Schreckenberg(1995)]{Nagel1992}
K.~Nagel and M.~Schreckenberg, \emph{J. Physique}, 1995, \textbf{51},
  2939\relax
\mciteBstWouldAddEndPuncttrue
\mciteSetBstMidEndSepPunct{\mcitedefaultmidpunct}
{\mcitedefaultendpunct}{\mcitedefaultseppunct}\relax
\EndOfBibitem
\bibitem[Umbanhowar \emph{et~al.}(1996)Umbanhowar, Melo, and
  Swinney]{Umbanhowar1996}
P.~B. Umbanhowar, F.~Melo and H.~L. Swinney, \emph{Nature}, 1996, \textbf{382},
  793\relax
\mciteBstWouldAddEndPuncttrue
\mciteSetBstMidEndSepPunct{\mcitedefaultmidpunct}
{\mcitedefaultendpunct}{\mcitedefaultseppunct}\relax
\EndOfBibitem
\bibitem[Jaeger and Nagel(1996)]{Jaeger1996}
H.~M. Jaeger and S.~Nagel, \emph{Rev. Mod. Phys.}, 1996, \textbf{68},
  1259\relax
\mciteBstWouldAddEndPuncttrue
\mciteSetBstMidEndSepPunct{\mcitedefaultmidpunct}
{\mcitedefaultendpunct}{\mcitedefaultseppunct}\relax
\EndOfBibitem
\bibitem[Aranson and Tsimring(2006)]{Aranson2006}
I.~S. Aranson and L.~S. Tsimring, \emph{Rev. Mod. Phys.}, 2006, \textbf{78},
  641\relax
\mciteBstWouldAddEndPuncttrue
\mciteSetBstMidEndSepPunct{\mcitedefaultmidpunct}
{\mcitedefaultendpunct}{\mcitedefaultseppunct}\relax
\EndOfBibitem
\bibitem[Werner(1995)]{Werner1995}
B.~T. Werner, \emph{Geology}, 1995, \textbf{23}, 1107\relax
\mciteBstWouldAddEndPuncttrue
\mciteSetBstMidEndSepPunct{\mcitedefaultmidpunct}
{\mcitedefaultendpunct}{\mcitedefaultseppunct}\relax
\EndOfBibitem
\bibitem[Sagan and Bagnold(1975)]{Sagan1975}
C.~Sagan and R.~A. Bagnold, \emph{Icarus}, 1975, \textbf{26}, 209\relax
\mciteBstWouldAddEndPuncttrue
\mciteSetBstMidEndSepPunct{\mcitedefaultmidpunct}
{\mcitedefaultendpunct}{\mcitedefaultseppunct}\relax
\EndOfBibitem
\bibitem[{Lorenz et al.}(2006)]{Lorenz2006}
R.~D. {Lorenz et al.}, \emph{Science}, 2006, \textbf{312}, 724\relax
\mciteBstWouldAddEndPuncttrue
\mciteSetBstMidEndSepPunct{\mcitedefaultmidpunct}
{\mcitedefaultendpunct}{\mcitedefaultseppunct}\relax
\EndOfBibitem
\bibitem[Brown \emph{et~al.}(2010)Brown, Rodenberg, Amend, Mozeika, Steltz,
  Zakin, Lipson, and Jaeger]{Brown2010}
E.~Brown, N.~Rodenberg, J.~Amend, A.~Mozeika, E.~Steltz, M.~R. Zakin, H.~Lipson
  and H.~M. Jaeger, \emph{PNAS}, 2010, \textbf{107}, 18809\relax
\mciteBstWouldAddEndPuncttrue
\mciteSetBstMidEndSepPunct{\mcitedefaultmidpunct}
{\mcitedefaultendpunct}{\mcitedefaultseppunct}\relax
\EndOfBibitem
\bibitem[Kakalios(2005)]{Kakalios2005}
J.~Kakalios, \emph{Am. J. Phys.}, 2005, \textbf{73}, 8\relax
\mciteBstWouldAddEndPuncttrue
\mciteSetBstMidEndSepPunct{\mcitedefaultmidpunct}
{\mcitedefaultendpunct}{\mcitedefaultseppunct}\relax
\EndOfBibitem
\bibitem[Frette \emph{et~al.}({1996})Frette, Christensen, Malthe-S\o{}renssen,
  Feder, J\o{}ssang, and Meakin]{Frette1996}
V.~Frette, K.~Christensen, A.~Malthe-S\o{}renssen, J.~Feder, T.~J\o{}ssang and
  P.~Meakin, \emph{{Nature}}, {1996}, \textbf{{379}}, {49--52}\relax
\mciteBstWouldAddEndPuncttrue
\mciteSetBstMidEndSepPunct{\mcitedefaultmidpunct}
{\mcitedefaultendpunct}{\mcitedefaultseppunct}\relax
\EndOfBibitem
\bibitem[CEGEO \emph{et~al.}(2012)CEGEO, Szarf, Voivret, Az\'ema, Richefeu,
  Delenne, Combe, Nouguier-Lehon, Villard, Sornay, Chaze, and Radja\"i]{CEGEO}
B.~S.-C. CEGEO, K.~Szarf, C.~Voivret, E.~Az\'ema, V.~Richefeu, J.-Y. Delenne,
  G.~Combe, C.~Nouguier-Lehon, P.~Villard, P.~Sornay, M.~Chaze and F.~Radja\"i,
  \emph{Europhys. Lett.}, 2012, \textbf{98}, 44008\relax
\mciteBstWouldAddEndPuncttrue
\mciteSetBstMidEndSepPunct{\mcitedefaultmidpunct}
{\mcitedefaultendpunct}{\mcitedefaultseppunct}\relax
\EndOfBibitem
\bibitem[Friedman and Robinson({2002})]{Friedman2002}
S.~Friedman and D.~Robinson, \emph{{Water Resources Res.}}, {2002},
  \textbf{{38}}, 18\relax
\mciteBstWouldAddEndPuncttrue
\mciteSetBstMidEndSepPunct{\mcitedefaultmidpunct}
{\mcitedefaultendpunct}{\mcitedefaultseppunct}\relax
\EndOfBibitem
\bibitem[Zuriguel \emph{et~al.}(2007)Zuriguel, Mullin, and
  Rotter]{Zuriguel2007}
I.~Zuriguel, T.~Mullin and J.~Rotter, \emph{Phys. Rev. Lett.}, 2007,
  \textbf{98}, 028001\relax
\mciteBstWouldAddEndPuncttrue
\mciteSetBstMidEndSepPunct{\mcitedefaultmidpunct}
{\mcitedefaultendpunct}{\mcitedefaultseppunct}\relax
\EndOfBibitem
\bibitem[Nakagawa \emph{et~al.}(1993)Nakagawa, Altobelli, Caprihan, Fukushima,
  and Jeong]{Nakagawa1993}
M.~Nakagawa, S.~A. Altobelli, A.~Caprihan, E.~Fukushima and E.-K. Jeong,
  \emph{Exp. Fluids}, 1993, \textbf{16}, 54\relax
\mciteBstWouldAddEndPuncttrue
\mciteSetBstMidEndSepPunct{\mcitedefaultmidpunct}
{\mcitedefaultendpunct}{\mcitedefaultseppunct}\relax
\EndOfBibitem
\bibitem[Sakaie \emph{et~al.}(2008)Sakaie, Fenistein, Carroll, van Hecke, and
  Umbanhowar]{Sakaie2008}
K.~Sakaie, D.~Fenistein, T.~Carroll, M.~van Hecke and P.~Umbanhowar,
  \emph{Europhys. Lett.}, 2008, \textbf{84}, 38001\relax
\mciteBstWouldAddEndPuncttrue
\mciteSetBstMidEndSepPunct{\mcitedefaultmidpunct}
{\mcitedefaultendpunct}{\mcitedefaultseppunct}\relax
\EndOfBibitem
\bibitem[Finger \emph{et~al.}(2006)Finger, Voigt, Stadler, Niessen, Naji, and
  Stannarius]{Finger2006}
T.~Finger, A.~Voigt, J.~Stadler, H.~Niessen, L.~Naji and R.~Stannarius,
  \emph{Phys. Rev. E}, 2006, \textbf{74}, 031312\relax
\mciteBstWouldAddEndPuncttrue
\mciteSetBstMidEndSepPunct{\mcitedefaultmidpunct}
{\mcitedefaultendpunct}{\mcitedefaultseppunct}\relax
\EndOfBibitem
\bibitem[Zhang \emph{et~al.}({2006})Zhang, Thompson, Reed, and
  Beenken]{Zhang2006}
W.~Zhang, K.~Thompson, A.~Reed and L.~Beenken, \emph{{Chem. Eng. Sci.}},
  {2006}, \textbf{{61}}, {8060}\relax
\mciteBstWouldAddEndPuncttrue
\mciteSetBstMidEndSepPunct{\mcitedefaultmidpunct}
{\mcitedefaultendpunct}{\mcitedefaultseppunct}\relax
\EndOfBibitem
\bibitem[Wegner \emph{et~al.}(2012)Wegner, B\"orzs\"onyi, Bien, Rose, and
  Stannarius]{Wegner2012}
S.~Wegner, T.~B\"orzs\"onyi, T.~Bien, G.~Rose and R.~Stannarius, \emph{Soft
  Matter}, 2012, \textbf{8}, 10950\relax
\mciteBstWouldAddEndPuncttrue
\mciteSetBstMidEndSepPunct{\mcitedefaultmidpunct}
{\mcitedefaultendpunct}{\mcitedefaultseppunct}\relax
\EndOfBibitem
\bibitem[B\"orzs\"onyi \emph{et~al.}(2012)B\"orzs\"onyi, Szab\'o, T\"or\"os,
  Wegner, T\"or\"ok, Somfai, Bien, and Stannarius]{Borzsonyi2012}
T.~B\"orzs\"onyi, B.~Szab\'o, G.~T\"or\"os, S.~Wegner, J.~T\"or\"ok, E.~Somfai,
  T.~Bien and R.~Stannarius, \emph{Phys. Rev. Lett.}, 2012, \textbf{108},
  228302\relax
\mciteBstWouldAddEndPuncttrue
\mciteSetBstMidEndSepPunct{\mcitedefaultmidpunct}
{\mcitedefaultendpunct}{\mcitedefaultseppunct}\relax
\EndOfBibitem
\bibitem[Ketcham and Carlson(2001)]{Ketcham2001}
R.~A. Ketcham and W.~D. Carlson, \emph{Computers \& Geosci.}, 2001,
  \textbf{27}, 381\relax
\mciteBstWouldAddEndPuncttrue
\mciteSetBstMidEndSepPunct{\mcitedefaultmidpunct}
{\mcitedefaultendpunct}{\mcitedefaultseppunct}\relax
\EndOfBibitem
\bibitem[Ehrichs \emph{et~al.}(1995)Ehrichs, Jaeger, Karczmar, Knight,
  Kuperman, and Nagel]{Ehrichs1995}
E.~E. Ehrichs, H.~M. Jaeger, G.~S. Karczmar, J.~B. Knight, V.~Y. Kuperman and
  S.~R. Nagel, \emph{Science}, 1995, \textbf{267}, 1632\relax
\mciteBstWouldAddEndPuncttrue
\mciteSetBstMidEndSepPunct{\mcitedefaultmidpunct}
{\mcitedefaultendpunct}{\mcitedefaultseppunct}\relax
\EndOfBibitem
\bibitem[Montillet and {Le Coq}(2001)]{Montillet2001}
A.~Montillet and L.~{Le Coq}, \emph{Powder Techn.}, 2001, \textbf{121},
  138\relax
\mciteBstWouldAddEndPuncttrue
\mciteSetBstMidEndSepPunct{\mcitedefaultmidpunct}
{\mcitedefaultendpunct}{\mcitedefaultseppunct}\relax
\EndOfBibitem
\bibitem[Howell \emph{et~al.}(1999)Howell, Behringer, and Veje]{Howell1999}
D.~Howell, R.~P. Behringer and C.~Veje, \emph{Phys. Rev. Lett.}, 1999,
  \textbf{82}, 5241\relax
\mciteBstWouldAddEndPuncttrue
\mciteSetBstMidEndSepPunct{\mcitedefaultmidpunct}
{\mcitedefaultendpunct}{\mcitedefaultseppunct}\relax
\EndOfBibitem
\bibitem[Daniels and Hayman(2008)]{Daniels2008}
K.~E. Daniels and N.~W. Hayman, \emph{J. Geophys. Res.}, 2008, \textbf{113},
  B11411\relax
\mciteBstWouldAddEndPuncttrue
\mciteSetBstMidEndSepPunct{\mcitedefaultmidpunct}
{\mcitedefaultendpunct}{\mcitedefaultseppunct}\relax
\EndOfBibitem
\bibitem[Buchalter and Bradley({1992})]{Buchalter1992}
B.~Buchalter and R.~Bradley, \emph{{Phys. Rev. A}}, {1992}, \textbf{{46}},
  {3046--3056}\relax
\mciteBstWouldAddEndPuncttrue
\mciteSetBstMidEndSepPunct{\mcitedefaultmidpunct}
{\mcitedefaultendpunct}{\mcitedefaultseppunct}\relax
\EndOfBibitem
\bibitem[Buchalter and Bradley({1994})]{Buchalter1994}
B.~Buchalter and R.~Bradley, \emph{{Europhys. Lett.}}, {1994}, \textbf{{26}},
  {159--164}\relax
\mciteBstWouldAddEndPuncttrue
\mciteSetBstMidEndSepPunct{\mcitedefaultmidpunct}
{\mcitedefaultendpunct}{\mcitedefaultseppunct}\relax
\EndOfBibitem
\bibitem[Stockely \emph{et~al.}({2003})Stockely, Diacou, and
  Franklin]{Stockely2003}
K.~Stockely, A.~Diacou and S.~Franklin, \emph{{Phys. Rev. E}}, {2003},
  \textbf{{67}}, {051302}\relax
\mciteBstWouldAddEndPuncttrue
\mciteSetBstMidEndSepPunct{\mcitedefaultmidpunct}
{\mcitedefaultendpunct}{\mcitedefaultseppunct}\relax
\EndOfBibitem
\bibitem[Moradi \emph{et~al.}({2010})Moradi, Hashemi, and
  Taghizadeh]{Moradi2010}
M.~Moradi, S.~Hashemi and F.~Taghizadeh, \emph{{Physica A}}, {2010},
  \textbf{{389}}, {4510--4519}\relax
\mciteBstWouldAddEndPuncttrue
\mciteSetBstMidEndSepPunct{\mcitedefaultmidpunct}
{\mcitedefaultendpunct}{\mcitedefaultseppunct}\relax
\EndOfBibitem
\bibitem[Aspelmeier \emph{et~al.}(1998)Aspelmeier, Giese, and
  Zippelius]{Aspelmeier1998}
T.~Aspelmeier, G.~Giese and A.~Zippelius, \emph{Phys. Rev. E}, 1998,
  \textbf{57}, 857\relax
\mciteBstWouldAddEndPuncttrue
\mciteSetBstMidEndSepPunct{\mcitedefaultmidpunct}
{\mcitedefaultendpunct}{\mcitedefaultseppunct}\relax
\EndOfBibitem
\bibitem[Cundall and Strack(1979)]{Cundall1979}
P.~A. Cundall and O.~D.~L. Strack, \emph{Geotechnique}, 1979, \textbf{29},
  47\relax
\mciteBstWouldAddEndPuncttrue
\mciteSetBstMidEndSepPunct{\mcitedefaultmidpunct}
{\mcitedefaultendpunct}{\mcitedefaultseppunct}\relax
\EndOfBibitem
\bibitem[Zhu \emph{et~al.}({2007})Zhu, Zhou, Yang, and Yu]{Zhu2007}
H.~P. Zhu, Z.~Y. Zhou, R.~Y. Yang and A.~B. Yu, \emph{{Chem. Eng. Sci.}},
  {2007}, \textbf{{62}}, {3378}\relax
\mciteBstWouldAddEndPuncttrue
\mciteSetBstMidEndSepPunct{\mcitedefaultmidpunct}
{\mcitedefaultendpunct}{\mcitedefaultseppunct}\relax
\EndOfBibitem
\bibitem[Moreau(1994)]{Moreau1994}
J.~J. Moreau, \emph{Eur. J. Mech., A/Solids}, 1994, \textbf{13}, 93\relax
\mciteBstWouldAddEndPuncttrue
\mciteSetBstMidEndSepPunct{\mcitedefaultmidpunct}
{\mcitedefaultendpunct}{\mcitedefaultseppunct}\relax
\EndOfBibitem
\bibitem[Jean(1999)]{Jean1999}
M.~Jean, \emph{Comput. Meth. Appl. Mech. Eng.}, 1999, \textbf{177}, 235\relax
\mciteBstWouldAddEndPuncttrue
\mciteSetBstMidEndSepPunct{\mcitedefaultmidpunct}
{\mcitedefaultendpunct}{\mcitedefaultseppunct}\relax
\EndOfBibitem
\bibitem[Donev \emph{et~al.}({2005})Donev, Torquato, and Stillinger]{Donev2005}
A.~Donev, S.~Torquato and F.~Stillinger, \emph{{J. Comput. Phys.}}, {2005},
  \textbf{{202}}, {765--793}\relax
\mciteBstWouldAddEndPuncttrue
\mciteSetBstMidEndSepPunct{\mcitedefaultmidpunct}
{\mcitedefaultendpunct}{\mcitedefaultseppunct}\relax
\EndOfBibitem
\bibitem[Hinrichsen \emph{et~al.}(1986)Hinrichsen, Feder, and
  J\o{}ssang]{Hinrichsen1986}
E.~L. Hinrichsen, J.~Feder and T.~J\o{}ssang, \emph{J. Stat. Phys.}, 1986,
  \textbf{44}, 793\relax
\mciteBstWouldAddEndPuncttrue
\mciteSetBstMidEndSepPunct{\mcitedefaultmidpunct}
{\mcitedefaultendpunct}{\mcitedefaultseppunct}\relax
\EndOfBibitem
\bibitem[Williams and Philipse({2003})]{Williams2003}
S.~Williams and A.~Philipse, \emph{{Phys. Rev. E}}, {2003}, \textbf{{67}},
  051301\relax
\mciteBstWouldAddEndPuncttrue
\mciteSetBstMidEndSepPunct{\mcitedefaultmidpunct}
{\mcitedefaultendpunct}{\mcitedefaultseppunct}\relax
\EndOfBibitem
\bibitem[Wouterse \emph{et~al.}({2009})Wouterse, Luding, and
  Philipse]{Wouterse2009}
A.~Wouterse, S.~Luding and A.~P. Philipse, \emph{{Granular Matter}}, {2009},
  \textbf{{11}}, {169--177}\relax
\mciteBstWouldAddEndPuncttrue
\mciteSetBstMidEndSepPunct{\mcitedefaultmidpunct}
{\mcitedefaultendpunct}{\mcitedefaultseppunct}\relax
\EndOfBibitem
\bibitem[Hinrichsen \emph{et~al.}(1990)Hinrichsen, Feder, and
  J\o{}ssang]{Hinrichsen1990}
E.~L. Hinrichsen, J.~Feder and T.~J\o{}ssang, \emph{Phys. Rev. A}, 1990,
  \textbf{41}, 4199\relax
\mciteBstWouldAddEndPuncttrue
\mciteSetBstMidEndSepPunct{\mcitedefaultmidpunct}
{\mcitedefaultendpunct}{\mcitedefaultseppunct}\relax
\EndOfBibitem
\bibitem[Torquato and Stillinger({2010})]{Torquato2010}
S.~Torquato and F.~H. Stillinger, \emph{{Rev. Mod. Phys.}}, {2010},
  \textbf{{82}}, {2633--2672}\relax
\mciteBstWouldAddEndPuncttrue
\mciteSetBstMidEndSepPunct{\mcitedefaultmidpunct}
{\mcitedefaultendpunct}{\mcitedefaultseppunct}\relax
\EndOfBibitem
\bibitem[M\"atzler({1997})]{Matzler1997}
C.~M\"atzler, \emph{{J. Appl. Phys.}}, {1997}, \textbf{{81}},
  {1509--1517}\relax
\mciteBstWouldAddEndPuncttrue
\mciteSetBstMidEndSepPunct{\mcitedefaultmidpunct}
{\mcitedefaultendpunct}{\mcitedefaultseppunct}\relax
\EndOfBibitem
\bibitem[Rankenburg and Zieve({2001})]{Rankenburg2001}
I.~Rankenburg and R.~Zieve, \emph{{Phys. Rev. E}}, {2001}, \textbf{{63}}, {art.
  no.--061303}\relax
\mciteBstWouldAddEndPuncttrue
\mciteSetBstMidEndSepPunct{\mcitedefaultmidpunct}
{\mcitedefaultendpunct}{\mcitedefaultseppunct}\relax
\EndOfBibitem
\bibitem[Mailman \emph{et~al.}({2009})Mailman, Schreck, O'Hern, and
  Chakraborty]{Mailman2009}
M.~Mailman, C.~F. Schreck, C.~S. O'Hern and B.~Chakraborty, \emph{{Phys. Rev.
  Lett.}}, {2009}, \textbf{{102}}, {255501}\relax
\mciteBstWouldAddEndPuncttrue
\mciteSetBstMidEndSepPunct{\mcitedefaultmidpunct}
{\mcitedefaultendpunct}{\mcitedefaultseppunct}\relax
\EndOfBibitem
\bibitem[Zeravcic \emph{et~al.}(2009)Zeravcic, Xu, Liu, Nagel, and van
  Saarloos]{Zeravcic2009}
Z.~Zeravcic, N.~Xu, A.~J. Liu, S.~R. Nagel and W.~van Saarloos, \emph{Europhys.
  Lett.}, 2009, \textbf{87}, 26001\relax
\mciteBstWouldAddEndPuncttrue
\mciteSetBstMidEndSepPunct{\mcitedefaultmidpunct}
{\mcitedefaultendpunct}{\mcitedefaultseppunct}\relax
\EndOfBibitem
\bibitem[Schreck \emph{et~al.}(2010)Schreck, Xu, and O'Hern]{Schreck2010}
C.~F. Schreck, N.~Xu and C.~S. O'Hern, \emph{Soft Matter}, 2010, \textbf{6},
  2960\relax
\mciteBstWouldAddEndPuncttrue
\mciteSetBstMidEndSepPunct{\mcitedefaultmidpunct}
{\mcitedefaultendpunct}{\mcitedefaultseppunct}\relax
\EndOfBibitem
\bibitem[Shen \emph{et~al.}(2012)Shen, Schreck, Chakraborty, Freed, and
  O'Hern]{Shen2012}
T.~Shen, C.~Schreck, B.~Chakraborty, D.~E. Freed and C.~S. O'Hern, \emph{Phys.
  Rev. E}, 2012, \textbf{86}, 041303\relax
\mciteBstWouldAddEndPuncttrue
\mciteSetBstMidEndSepPunct{\mcitedefaultmidpunct}
{\mcitedefaultendpunct}{\mcitedefaultseppunct}\relax
\EndOfBibitem
\bibitem[Olson \emph{et~al.}({2002})Olson, Reichhardt, McCloskey, and
  Zieve]{Olson2002}
C.~Olson, C.~Reichhardt, M.~McCloskey and R.~Zieve, \emph{{Europhys. Lett.}},
  {2002}, \textbf{{57}}, {904--910}\relax
\mciteBstWouldAddEndPuncttrue
\mciteSetBstMidEndSepPunct{\mcitedefaultmidpunct}
{\mcitedefaultendpunct}{\mcitedefaultseppunct}\relax
\EndOfBibitem
\bibitem[Donev \emph{et~al.}({2004})Donev, Cisse, Sachs, Variano, Stillinger,
  Connelly, Torquato, and Chaikin]{Donev2004}
A.~Donev, I.~Cisse, D.~Sachs, E.~Variano, F.~Stillinger, R.~Connelly,
  S.~Torquato and P.~Chaikin, \emph{{Science}}, {2004}, \textbf{{303}},
  {990--993}\relax
\mciteBstWouldAddEndPuncttrue
\mciteSetBstMidEndSepPunct{\mcitedefaultmidpunct}
{\mcitedefaultendpunct}{\mcitedefaultseppunct}\relax
\EndOfBibitem
\bibitem[Baram and Lind({2012})]{Baram2012}
R.~M. Baram and P.~G. Lind, \emph{{Phys. Rev. E}}, {2012}, \textbf{{85}},
  041301\relax
\mciteBstWouldAddEndPuncttrue
\mciteSetBstMidEndSepPunct{\mcitedefaultmidpunct}
{\mcitedefaultendpunct}{\mcitedefaultseppunct}\relax
\EndOfBibitem
\bibitem[Chaikin \emph{et~al.}(2006)Chaikin, Donev, Man, Stillinger, and
  Torquato]{Chaikin2006}
P.~M. Chaikin, A.~Donev, W.~Man, F.~H. Stillinger and S.~Torquato, \emph{Ind.
  Eng. Chem. Res.}, 2006, \textbf{45}, 6960\relax
\mciteBstWouldAddEndPuncttrue
\mciteSetBstMidEndSepPunct{\mcitedefaultmidpunct}
{\mcitedefaultendpunct}{\mcitedefaultseppunct}\relax
\EndOfBibitem
\bibitem[Donev \emph{et~al.}({2007})Donev, Connelly, Stillinger, and
  Torquato]{Donev2007}
A.~Donev, R.~Connelly, F.~H. Stillinger and S.~Torquato, \emph{{Phys. Rev. E}},
  {2007}, \textbf{{75}}, 051304\relax
\mciteBstWouldAddEndPuncttrue
\mciteSetBstMidEndSepPunct{\mcitedefaultmidpunct}
{\mcitedefaultendpunct}{\mcitedefaultseppunct}\relax
\EndOfBibitem
\bibitem[Sacanna \emph{et~al.}({2007})Sacanna, Rossi, Wouterse, and
  Philipse]{Sacanna2007}
S.~Sacanna, L.~Rossi, A.~Wouterse and A.~P. Philipse, \emph{{J. Phys. - Cond.
  Mat.}}, {2007}, \textbf{{19}}, 376108\relax
\mciteBstWouldAddEndPuncttrue
\mciteSetBstMidEndSepPunct{\mcitedefaultmidpunct}
{\mcitedefaultendpunct}{\mcitedefaultseppunct}\relax
\EndOfBibitem
\bibitem[Philipse({1996})]{Philipse1996}
A.~P. Philipse, \emph{{Langmuir}}, {1996}, \textbf{{12}}, {1127}\relax
\mciteBstWouldAddEndPuncttrue
\mciteSetBstMidEndSepPunct{\mcitedefaultmidpunct}
{\mcitedefaultendpunct}{\mcitedefaultseppunct}\relax
\EndOfBibitem
\bibitem[Philipse({1996})]{Philipse1996c}
A.~P. Philipse, \emph{{Langmuir}}, {1996}, \textbf{{12}}, {5971}\relax
\mciteBstWouldAddEndPuncttrue
\mciteSetBstMidEndSepPunct{\mcitedefaultmidpunct}
{\mcitedefaultendpunct}{\mcitedefaultseppunct}\relax
\EndOfBibitem
\bibitem[Novellani \emph{et~al.}({2000})Novellani, Santini, and
  Tadrist]{Novellani2000}
M.~Novellani, R.~Santini and L.~Tadrist, \emph{{Eur. Phys. B}}, {2000},
  \textbf{{13}}, {571}\relax
\mciteBstWouldAddEndPuncttrue
\mciteSetBstMidEndSepPunct{\mcitedefaultmidpunct}
{\mcitedefaultendpunct}{\mcitedefaultseppunct}\relax
\EndOfBibitem
\bibitem[Blouwolff and Fraden(2006)]{Blouwolff2006}
J.~Blouwolff and S.~Fraden, \emph{Europhys. Lett.}, 2006, \textbf{76},
  1095\relax
\mciteBstWouldAddEndPuncttrue
\mciteSetBstMidEndSepPunct{\mcitedefaultmidpunct}
{\mcitedefaultendpunct}{\mcitedefaultseppunct}\relax
\EndOfBibitem
\bibitem[Milevski(1978)]{Milevski1978}
J.~V. Milevski, \emph{Ind. Eng. Chem. Prod. Res.}, 1978, \textbf{17}, 363\relax
\mciteBstWouldAddEndPuncttrue
\mciteSetBstMidEndSepPunct{\mcitedefaultmidpunct}
{\mcitedefaultendpunct}{\mcitedefaultseppunct}\relax
\EndOfBibitem
\bibitem[Nardin \emph{et~al.}(1985)Nardin, Papirer, and Schultz]{Nardin1985}
M.~Nardin, E.~Papirer and J.~Schultz, \emph{Powder Techn.}, 1985, \textbf{44},
  131\relax
\mciteBstWouldAddEndPuncttrue
\mciteSetBstMidEndSepPunct{\mcitedefaultmidpunct}
{\mcitedefaultendpunct}{\mcitedefaultseppunct}\relax
\EndOfBibitem
\bibitem[Kyrylyuk \emph{et~al.}({2011})Kyrylyuk, van~de Haar, Rossi, Wouterse,
  and Philipse]{Kyrylyuk2011b}
A.~V. Kyrylyuk, M.~A. van~de Haar, L.~Rossi, A.~Wouterse and A.~P. Philipse,
  \emph{{Soft Matter}}, {2011}, \textbf{{7}}, {1671--1674}\relax
\mciteBstWouldAddEndPuncttrue
\mciteSetBstMidEndSepPunct{\mcitedefaultmidpunct}
{\mcitedefaultendpunct}{\mcitedefaultseppunct}\relax
\EndOfBibitem
\bibitem[Kyrylyuk and Philipse({2011})]{Kyrylyuk2011}
A.~V. Kyrylyuk and A.~P. Philipse, \emph{{Phys. Stat. Sol. A}}, {2011},
  \textbf{{208}}, {2299--2302}\relax
\mciteBstWouldAddEndPuncttrue
\mciteSetBstMidEndSepPunct{\mcitedefaultmidpunct}
{\mcitedefaultendpunct}{\mcitedefaultseppunct}\relax
\EndOfBibitem
\bibitem[Villarruel \emph{et~al.}({2000})Villarruel, Lauderdale, Mueth, and
  Jaeger]{Villarruel2000}
F.~Villarruel, B.~Lauderdale, D.~Mueth and H.~Jaeger, \emph{{Phys. Rev. E}},
  {2000}, \textbf{{61}}, {6914--6921}\relax
\mciteBstWouldAddEndPuncttrue
\mciteSetBstMidEndSepPunct{\mcitedefaultmidpunct}
{\mcitedefaultendpunct}{\mcitedefaultseppunct}\relax
\EndOfBibitem
\bibitem[Lumay and Vandewalle({2004})]{Lumay2004}
G.~Lumay and N.~Vandewalle, \emph{{Phys. Rev. E}}, {2004}, \textbf{{70}},
  {051314}\relax
\mciteBstWouldAddEndPuncttrue
\mciteSetBstMidEndSepPunct{\mcitedefaultmidpunct}
{\mcitedefaultendpunct}{\mcitedefaultseppunct}\relax
\EndOfBibitem
\bibitem[Lumay and Vandewalle(2006)]{Lumay2006}
G.~Lumay and N.~Vandewalle, \emph{{Phys. Rev. E}}, 2006, \textbf{74},
  {021301}\relax
\mciteBstWouldAddEndPuncttrue
\mciteSetBstMidEndSepPunct{\mcitedefaultmidpunct}
{\mcitedefaultendpunct}{\mcitedefaultseppunct}\relax
\EndOfBibitem
\bibitem[Vandewalle \emph{et~al.}(2007)Vandewalle, Lumay, Gerasimov, and
  Ludewig]{Vandewalle2007}
N.~Vandewalle, G.~Lumay, O.~Gerasimov and F.~Ludewig, \emph{Eur. Phys. J. E},
  2007, \textbf{22}, 241\relax
\mciteBstWouldAddEndPuncttrue
\mciteSetBstMidEndSepPunct{\mcitedefaultmidpunct}
{\mcitedefaultendpunct}{\mcitedefaultseppunct}\relax
\EndOfBibitem
\bibitem[{Cruz Hidalgo} \emph{et~al.}({2009}){Cruz Hidalgo}, Zuriguel, Maza,
  and Pagonabarraga]{CruzHidalgo2009}
R.~{Cruz Hidalgo}, I.~Zuriguel, D.~Maza and I.~Pagonabarraga, \emph{{Phys. Rev.
  Lett.}}, {2009}, \textbf{{103}}, 118001\relax
\mciteBstWouldAddEndPuncttrue
\mciteSetBstMidEndSepPunct{\mcitedefaultmidpunct}
{\mcitedefaultendpunct}{\mcitedefaultseppunct}\relax
\EndOfBibitem
\bibitem[{Cruz Hidalgo} \emph{et~al.}({2010}){Cruz Hidalgo}, Zuriguel, Maza,
  and Pagonabarraga]{CruzHidalgo2010}
R.~{Cruz Hidalgo}, I.~Zuriguel, D.~Maza and I.~Pagonabarraga, \emph{J. Stat.
  Mech.-Theor. Exp.}, {2010},  P06025\relax
\mciteBstWouldAddEndPuncttrue
\mciteSetBstMidEndSepPunct{\mcitedefaultmidpunct}
{\mcitedefaultendpunct}{\mcitedefaultseppunct}\relax
\EndOfBibitem
\bibitem[Kanzaki \emph{et~al.}({2011})Kanzaki, Acevedo, Zuriguel,
  Pagonabarraga, Maza, and Hidalgo]{Kanzaki2011}
T.~Kanzaki, M.~Acevedo, I.~Zuriguel, I.~Pagonabarraga, D.~Maza and R.~C.
  Hidalgo, \emph{{Eur. Phys. J. E}}, {2011}, \textbf{{34}}, 133\relax
\mciteBstWouldAddEndPuncttrue
\mciteSetBstMidEndSepPunct{\mcitedefaultmidpunct}
{\mcitedefaultendpunct}{\mcitedefaultseppunct}\relax
\EndOfBibitem
\bibitem[Hidalgo \emph{et~al.}({2012})Hidalgo, Kadau, Kanzaki, and
  Herrmann]{Hidalgo2012}
R.~C. Hidalgo, D.~Kadau, T.~Kanzaki and H.~J. Herrmann, \emph{{Granular
  Matter}}, {2012}, \textbf{{14}}, {191--196}\relax
\mciteBstWouldAddEndPuncttrue
\mciteSetBstMidEndSepPunct{\mcitedefaultmidpunct}
{\mcitedefaultendpunct}{\mcitedefaultseppunct}\relax
\EndOfBibitem
\bibitem[Az\'ema and Radja\"i({2010})]{Azema2010}
E.~Az\'ema and F.~Radja\"i, \emph{{Phys. Rev. E}}, {2010}, \textbf{{81}},
  051304\relax
\mciteBstWouldAddEndPuncttrue
\mciteSetBstMidEndSepPunct{\mcitedefaultmidpunct}
{\mcitedefaultendpunct}{\mcitedefaultseppunct}\relax
\EndOfBibitem
\bibitem[Az\'ema and Radja\"i(2012)]{Azema2012}
E.~Az\'ema and F.~Radja\"i, \emph{{Phys. Rev. E}}, 2012, \textbf{85},
  031303\relax
\mciteBstWouldAddEndPuncttrue
\mciteSetBstMidEndSepPunct{\mcitedefaultmidpunct}
{\mcitedefaultendpunct}{\mcitedefaultseppunct}\relax
\EndOfBibitem
\bibitem[Nouguier-Lehon \emph{et~al.}({2003})Nouguier-Lehon, Cambou, and
  Vincens]{Nouguier-Lehon2003}
C.~Nouguier-Lehon, B.~Cambou and E.~Vincens, \emph{Int. J. Numer. Anal. Meth.
  Geomechanics}, {2003}, \textbf{{27}}, {1207--1226}\relax
\mciteBstWouldAddEndPuncttrue
\mciteSetBstMidEndSepPunct{\mcitedefaultmidpunct}
{\mcitedefaultendpunct}{\mcitedefaultseppunct}\relax
\EndOfBibitem
\bibitem[Nouguier-Lehon({2010})]{Nouguier-Lehon2010}
C.~Nouguier-Lehon, \emph{{Compt. Rend. Mecanique}}, {2010}, \textbf{{338}},
  {587--595}\relax
\mciteBstWouldAddEndPuncttrue
\mciteSetBstMidEndSepPunct{\mcitedefaultmidpunct}
{\mcitedefaultendpunct}{\mcitedefaultseppunct}\relax
\EndOfBibitem
\bibitem[Wiacek \emph{et~al.}({2012})Wiacek, Molenda, Horabik, and
  Ooi]{Wiacek2012}
J.~Wiacek, M.~Molenda, J.~Horabik and J.~Y. Ooi, \emph{{Powder Techn.}},
  {2012}, \textbf{{217}}, {435--442}\relax
\mciteBstWouldAddEndPuncttrue
\mciteSetBstMidEndSepPunct{\mcitedefaultmidpunct}
{\mcitedefaultendpunct}{\mcitedefaultseppunct}\relax
\EndOfBibitem
\bibitem[Boton \emph{et~al.}(2013)Boton, Az\'ema, Estrada, Radja\"i, and
  Lizcano]{Boton2013}
M.~Boton, E.~Az\'ema, N.~Estrada, F.~Radja\"i and A.~Lizcano, \emph{{Phys. Rev.
  E}}, 2013, \textbf{87}, 032206\relax
\mciteBstWouldAddEndPuncttrue
\mciteSetBstMidEndSepPunct{\mcitedefaultmidpunct}
{\mcitedefaultendpunct}{\mcitedefaultseppunct}\relax
\EndOfBibitem
\bibitem[Ng({2004})]{Ng2004}
T.~Ng, \emph{{J. Geotechn. Geoenv. Eng.}}, {2004}, \textbf{{130}},
  {1077--1083}\relax
\mciteBstWouldAddEndPuncttrue
\mciteSetBstMidEndSepPunct{\mcitedefaultmidpunct}
{\mcitedefaultendpunct}{\mcitedefaultseppunct}\relax
\EndOfBibitem
\bibitem[Ng({2009})]{Ng2009}
T.-T. Ng, \emph{{Mech. Materials}}, {2009}, \textbf{{41}}, {748--763}\relax
\mciteBstWouldAddEndPuncttrue
\mciteSetBstMidEndSepPunct{\mcitedefaultmidpunct}
{\mcitedefaultendpunct}{\mcitedefaultseppunct}\relax
\EndOfBibitem
\bibitem[Ng({2009})]{Ng2009b}
T.-T. Ng, \emph{{Int. J. Numer. Anal. Meth. Geomechanics}}, {2009},
  \textbf{{33}}, {511--527}\relax
\mciteBstWouldAddEndPuncttrue
\mciteSetBstMidEndSepPunct{\mcitedefaultmidpunct}
{\mcitedefaultendpunct}{\mcitedefaultseppunct}\relax
\EndOfBibitem
\bibitem[Antony and Kuhn(2004)]{Antony2004}
S.~J. Antony and M.~R. Kuhn, \emph{International Journal of Solids and
  Structures}, 2004, \textbf{41}, 5863\relax
\mciteBstWouldAddEndPuncttrue
\mciteSetBstMidEndSepPunct{\mcitedefaultmidpunct}
{\mcitedefaultendpunct}{\mcitedefaultseppunct}\relax
\EndOfBibitem
\bibitem[Trepanier and Franklin({2010})]{Trepanier2010}
M.~Trepanier and S.~V. Franklin, \emph{{Phys. Rev. E}}, {2010}, \textbf{{82}},
  011308\relax
\mciteBstWouldAddEndPuncttrue
\mciteSetBstMidEndSepPunct{\mcitedefaultmidpunct}
{\mcitedefaultendpunct}{\mcitedefaultseppunct}\relax
\EndOfBibitem
\bibitem[Tapia-McClung and Zenit({2012})]{TapiaMcClung2012}
H.~Tapia-McClung and R.~Zenit, \emph{{Phys. Rev. E}}, {2012}, \textbf{{85}},
  {061304}\relax
\mciteBstWouldAddEndPuncttrue
\mciteSetBstMidEndSepPunct{\mcitedefaultmidpunct}
{\mcitedefaultendpunct}{\mcitedefaultseppunct}\relax
\EndOfBibitem
\bibitem[Zhou \emph{et~al.}({2011})Zhou, Pinson, Zou, and Yu]{Zhou2011}
Z.~Y. Zhou, D.~Pinson, R.~P. Zou and A.~B. Yu, \emph{{Chem. Eng. Sci.}},
  {2011}, \textbf{{66}}, {6128--6145}\relax
\mciteBstWouldAddEndPuncttrue
\mciteSetBstMidEndSepPunct{\mcitedefaultmidpunct}
{\mcitedefaultendpunct}{\mcitedefaultseppunct}\relax
\EndOfBibitem
\bibitem[Narayan \emph{et~al.}({2006})Narayan, Menon, and
  Ramaswamy]{Narayan2006}
V.~Narayan, N.~Menon and S.~Ramaswamy, \emph{{J. Stat. Mech.}}, {2006},
  {P01005}\relax
\mciteBstWouldAddEndPuncttrue
\mciteSetBstMidEndSepPunct{\mcitedefaultmidpunct}
{\mcitedefaultendpunct}{\mcitedefaultseppunct}\relax
\EndOfBibitem
\bibitem[Ben-Naim and Krapivsky({2006})]{BenNaim2006}
E.~Ben-Naim and P.~L. Krapivsky, \emph{{Phys. Rev. E}}, {2006}, \textbf{{73}},
  {031109}\relax
\mciteBstWouldAddEndPuncttrue
\mciteSetBstMidEndSepPunct{\mcitedefaultmidpunct}
{\mcitedefaultendpunct}{\mcitedefaultseppunct}\relax
\EndOfBibitem
\bibitem[Galanis \emph{et~al.}({2006})Galanis, Harries, Sackett, Losert, and
  Nossal]{Galanis2006}
J.~Galanis, D.~Harries, D.~L. Sackett, W.~Losert and R.~Nossal, \emph{{Phys.
  Rev. Lett.}}, {2006}, \textbf{{96}}, {028002}\relax
\mciteBstWouldAddEndPuncttrue
\mciteSetBstMidEndSepPunct{\mcitedefaultmidpunct}
{\mcitedefaultendpunct}{\mcitedefaultseppunct}\relax
\EndOfBibitem
\bibitem[Galanis \emph{et~al.}({2010})Galanis, Nossal, Losert, and
  Harries]{Galanis2010}
J.~Galanis, R.~Nossal, W.~Losert and D.~Harries, \emph{{Phys. Rev. Lett.}},
  {2010}, \textbf{{105}}, {168001}\relax
\mciteBstWouldAddEndPuncttrue
\mciteSetBstMidEndSepPunct{\mcitedefaultmidpunct}
{\mcitedefaultendpunct}{\mcitedefaultseppunct}\relax
\EndOfBibitem
\bibitem[Aranson \emph{et~al.}(2007)Aranson, Volfson, and
  Tsimring]{Aranson2007}
I.~S. Aranson, D.~Volfson and L.~S. Tsimring, \emph{Phys. Rev. E}, 2007,
  \textbf{75}, 051301\relax
\mciteBstWouldAddEndPuncttrue
\mciteSetBstMidEndSepPunct{\mcitedefaultmidpunct}
{\mcitedefaultendpunct}{\mcitedefaultseppunct}\relax
\EndOfBibitem
\bibitem[Blair \emph{et~al.}(2003)Blair, Neicu, and Kudrolli]{Blair2003}
D.~L. Blair, T.~Neicu and A.~Kudrolli, \emph{Phys. Rev. E}, 2003, \textbf{67},
  031303\relax
\mciteBstWouldAddEndPuncttrue
\mciteSetBstMidEndSepPunct{\mcitedefaultmidpunct}
{\mcitedefaultendpunct}{\mcitedefaultseppunct}\relax
\EndOfBibitem
\bibitem[Volfson \emph{et~al.}(2004)Volfson, Kudrolli, and
  Tsimring]{Volfson2004}
D.~Volfson, A.~Kudrolli and L.~S. Tsimring, \emph{Phys. Rev. E}, 2004,
  \textbf{70}, 051312\relax
\mciteBstWouldAddEndPuncttrue
\mciteSetBstMidEndSepPunct{\mcitedefaultmidpunct}
{\mcitedefaultendpunct}{\mcitedefaultseppunct}\relax
\EndOfBibitem
\bibitem[Wambaugh \emph{et~al.}({2002})Wambaugh, Reichhardt, and
  Olson]{Wambaugh2002}
J.~Wambaugh, C.~Reichhardt and C.~Olson, \emph{{Phys. Rev. E}}, {2002},
  \textbf{{65}}, 031308\relax
\mciteBstWouldAddEndPuncttrue
\mciteSetBstMidEndSepPunct{\mcitedefaultmidpunct}
{\mcitedefaultendpunct}{\mcitedefaultseppunct}\relax
\EndOfBibitem
\bibitem[Narayan \emph{et~al.}({2007})Narayan, Ramaswamy, and
  Menon]{Narayan2007}
V.~Narayan, S.~Ramaswamy and N.~Menon, \emph{{Science}}, {2007},
  \textbf{{317}}, {105}\relax
\mciteBstWouldAddEndPuncttrue
\mciteSetBstMidEndSepPunct{\mcitedefaultmidpunct}
{\mcitedefaultendpunct}{\mcitedefaultseppunct}\relax
\EndOfBibitem
\bibitem[Aranson \emph{et~al.}(2008)Aranson, Snezhko, Olafsen, and
  Urbach]{Aranson2008}
I.~S. Aranson, A.~Snezhko, J.~S. Olafsen and J.~S. Urbach, \emph{Science},
  2008, \textbf{320}, 612c\relax
\mciteBstWouldAddEndPuncttrue
\mciteSetBstMidEndSepPunct{\mcitedefaultmidpunct}
{\mcitedefaultendpunct}{\mcitedefaultseppunct}\relax
\EndOfBibitem
\bibitem[Deseigne \emph{et~al.}(2010)Deseigne, Dauchot, and
  Chat\'e]{Deseigne2010}
J.~Deseigne, O.~Dauchot and H.~Chat\'e, \emph{Phys. Rev. Lett.}, 2010,
  \textbf{105}, 098001\relax
\mciteBstWouldAddEndPuncttrue
\mciteSetBstMidEndSepPunct{\mcitedefaultmidpunct}
{\mcitedefaultendpunct}{\mcitedefaultseppunct}\relax
\EndOfBibitem
\bibitem[Ramaswamy \emph{et~al.}(2003)Ramaswamy, Simha, and
  Toner]{Ramaswamy2003}
S.~Ramaswamy, R.~Simha and J.~Toner, \emph{Europhys. Lett.}, 2003, \textbf{62},
  196\relax
\mciteBstWouldAddEndPuncttrue
\mciteSetBstMidEndSepPunct{\mcitedefaultmidpunct}
{\mcitedefaultendpunct}{\mcitedefaultseppunct}\relax
\EndOfBibitem
\bibitem[H.~Chat\'e and Montagne(2006)]{Chate2006}
F.~G. H.~Chat\'e and R.~Montagne, \emph{Phys. Rev. Lett.}, 2006, \textbf{96},
  180602\relax
\mciteBstWouldAddEndPuncttrue
\mciteSetBstMidEndSepPunct{\mcitedefaultmidpunct}
{\mcitedefaultendpunct}{\mcitedefaultseppunct}\relax
\EndOfBibitem
\bibitem[Kudrolli \emph{et~al.}({2008})Kudrolli, Lumay, Volfson, and
  Tsimring]{Kudrolli2008}
A.~Kudrolli, G.~Lumay, D.~Volfson and L.~S. Tsimring, \emph{{Phys. Rev.
  Lett.}}, {2008}, \textbf{{100}}, {058001}\relax
\mciteBstWouldAddEndPuncttrue
\mciteSetBstMidEndSepPunct{\mcitedefaultmidpunct}
{\mcitedefaultendpunct}{\mcitedefaultseppunct}\relax
\EndOfBibitem
\bibitem[Kudrolli({2010})]{Kudrolli2010}
A.~Kudrolli, \emph{{Phys. Rev. Lett.}}, {2010}, \textbf{{104}}, {088001}\relax
\mciteBstWouldAddEndPuncttrue
\mciteSetBstMidEndSepPunct{\mcitedefaultmidpunct}
{\mcitedefaultendpunct}{\mcitedefaultseppunct}\relax
\EndOfBibitem
\bibitem[Ramaioli \emph{et~al.}(2005)Ramaioli, Pournin, and
  Liebling]{Ramaioli2005}
M.~Ramaioli, L.~Pournin and T.~M. Liebling, in \emph{Proc. Powder and Grains
  Conference, Stuttgart}, ed. R.~García-Rojo, H.~J. Herrmann and S.~McNamara,
  A. A. Balkema Publishers, Leiden, 2005, vol.~2, p. 1359\relax
\mciteBstWouldAddEndPuncttrue
\mciteSetBstMidEndSepPunct{\mcitedefaultmidpunct}
{\mcitedefaultendpunct}{\mcitedefaultseppunct}\relax
\EndOfBibitem
\bibitem[Ramaioli \emph{et~al.}({2007})Ramaioli, Pournin, and
  Liebling]{Ramaioli2007}
M.~Ramaioli, L.~Pournin and T.~M. Liebling, \emph{{Phys. Rev. E}}, {2007},
  \textbf{{76}}, 021304\relax
\mciteBstWouldAddEndPuncttrue
\mciteSetBstMidEndSepPunct{\mcitedefaultmidpunct}
{\mcitedefaultendpunct}{\mcitedefaultseppunct}\relax
\EndOfBibitem
\bibitem[Ribiere \emph{et~al.}({2005})Ribiere, Richard, Dallnay, and
  Bideau]{Ribiere2005}
P.~Ribiere, P.~Richard, R.~Dallnay and D.~Bideau, \emph{{Phys. Rev. E}},
  {2005}, \textbf{{71}}, {011304}\relax
\mciteBstWouldAddEndPuncttrue
\mciteSetBstMidEndSepPunct{\mcitedefaultmidpunct}
{\mcitedefaultendpunct}{\mcitedefaultseppunct}\relax
\EndOfBibitem
\bibitem[Caulkin \emph{et~al.}({2010})Caulkin, Jia, Fairweather, and
  Williams]{Caulkin2010}
R.~Caulkin, X.~Jia, M.~Fairweather and R.~A. Williams, \emph{{Phys. Rev. E}},
  {2010}, \textbf{{81}}, 051302\relax
\mciteBstWouldAddEndPuncttrue
\mciteSetBstMidEndSepPunct{\mcitedefaultmidpunct}
{\mcitedefaultendpunct}{\mcitedefaultseppunct}\relax
\EndOfBibitem
\bibitem[Baxter and Behringer(1990)]{Baxter1990}
G.~W. Baxter and R.~P. Behringer, \emph{Phys. Rev. A}, 1990, \textbf{42},
  1017\relax
\mciteBstWouldAddEndPuncttrue
\mciteSetBstMidEndSepPunct{\mcitedefaultmidpunct}
{\mcitedefaultendpunct}{\mcitedefaultseppunct}\relax
\EndOfBibitem
\bibitem[Cleary(1999)]{Cleary1999}
P.~W. Cleary, \emph{Proc. 2nd Int. Conf. on CFD in the Minerals and Process
  Ind. CSIRO, Melbourne, Australia}, 1999,  71\relax
\mciteBstWouldAddEndPuncttrue
\mciteSetBstMidEndSepPunct{\mcitedefaultmidpunct}
{\mcitedefaultendpunct}{\mcitedefaultseppunct}\relax
\EndOfBibitem
\bibitem[Cleary and Sawley({2002})]{Cleary2002}
P.~Cleary and M.~Sawley, \emph{{Appl. Math. Modelling}}, {2002}, \textbf{{26}},
  {89--111}\relax
\mciteBstWouldAddEndPuncttrue
\mciteSetBstMidEndSepPunct{\mcitedefaultmidpunct}
{\mcitedefaultendpunct}{\mcitedefaultseppunct}\relax
\EndOfBibitem
\bibitem[Sielamowicz \emph{et~al.}(2005)Sielamowicz, Blonski, and
  Kowalewski]{Sielamowicz2005}
I.~Sielamowicz, S.~Blonski and T.~Kowalewski, \emph{Chemical Engineering
  Science}, 2005, \textbf{60}, 589--598\relax
\mciteBstWouldAddEndPuncttrue
\mciteSetBstMidEndSepPunct{\mcitedefaultmidpunct}
{\mcitedefaultendpunct}{\mcitedefaultseppunct}\relax
\EndOfBibitem
\bibitem[Jin \emph{et~al.}({2010})Jin, Tao, and Zhong]{Jin2010}
B.~Jin, H.~Tao and W.~Zhong, \emph{{Chinese J. Chem. Eng.}}, {2010},
  \textbf{{18}}, {931--939}\relax
\mciteBstWouldAddEndPuncttrue
\mciteSetBstMidEndSepPunct{\mcitedefaultmidpunct}
{\mcitedefaultendpunct}{\mcitedefaultseppunct}\relax
\EndOfBibitem
\bibitem[Tao \emph{et~al.}(2010)Tao, Jin, Zhong, Ren, Zhang, and Xiao]{Tao2010}
H.~Tao, B.~Jin, W.~Zhong, B.~Ren, Y.~Zhang and R.~Xiao, \emph{Chem. Eng. Proc.:
  Processing Intensification}, 2010, \textbf{49}, 151\relax
\mciteBstWouldAddEndPuncttrue
\mciteSetBstMidEndSepPunct{\mcitedefaultmidpunct}
{\mcitedefaultendpunct}{\mcitedefaultseppunct}\relax
\EndOfBibitem
\bibitem[Langston \emph{et~al.}(2004)Langston, Al-Awamleh, Fraige, and
  Asmar]{Langston2004}
P.~A. Langston, M.~A. Al-Awamleh, F.~Y. Fraige and B.~N. Asmar, \emph{Chem.
  Eng. Sci.}, 2004, \textbf{59}, 425\relax
\mciteBstWouldAddEndPuncttrue
\mciteSetBstMidEndSepPunct{\mcitedefaultmidpunct}
{\mcitedefaultendpunct}{\mcitedefaultseppunct}\relax
\EndOfBibitem
\bibitem[Li \emph{et~al.}(2004)Li, Langston, Webb, and Dyakowski]{Li2004}
J.~Li, P.~A. Langston, C.~Webb and T.~Dyakowski, \emph{Chem. Eng. Sci.}, 2004,
  \textbf{59}, 5917\relax
\mciteBstWouldAddEndPuncttrue
\mciteSetBstMidEndSepPunct{\mcitedefaultmidpunct}
{\mcitedefaultendpunct}{\mcitedefaultseppunct}\relax
\EndOfBibitem
\bibitem[Markauskas and Ka\v{c}ianauskas({2011})]{Markauskas2011}
D.~Markauskas and R.~Ka\v{c}ianauskas \emph{{Granular Matter}}, {2011},
  \textbf{{13}}, {143--148}\relax
\mciteBstWouldAddEndPuncttrue
\mciteSetBstMidEndSepPunct{\mcitedefaultmidpunct}
{\mcitedefaultendpunct}{\mcitedefaultseppunct}\relax
\EndOfBibitem
\bibitem[Zuriguel \emph{et~al.}({2005})Zuriguel, Garcimart\'in, Maza,
  Pugnaloni, and Pastor]{Zuriguel2005}
I.~Zuriguel, A.~Garcimart\'in, D.~Maza, L.~A. Pugnaloni and J.~M. Pastor,
  \emph{{Phys. Rev. E}}, {2005}, \textbf{{71}}, 051303\relax
\mciteBstWouldAddEndPuncttrue
\mciteSetBstMidEndSepPunct{\mcitedefaultmidpunct}
{\mcitedefaultendpunct}{\mcitedefaultseppunct}\relax
\EndOfBibitem
\bibitem[Gleiser({2001})]{Gleiser2001}
P.~Gleiser, \emph{{Physica A}}, {2001}, \textbf{{295}}, {311--315}\relax
\mciteBstWouldAddEndPuncttrue
\mciteSetBstMidEndSepPunct{\mcitedefaultmidpunct}
{\mcitedefaultendpunct}{\mcitedefaultseppunct}\relax
\EndOfBibitem
\bibitem[Ahlgren \emph{et~al.}({2002})Ahlgren, Avlund, Klewe, Pedersen, and
  Corral]{Ahlgren2002}
P.~Ahlgren, M.~Avlund, I.~Klewe, J.~Pedersen and A.~Corral, \emph{{Phys. Rev.
  E}}, {2002}, \textbf{{66}}, 031305\relax
\mciteBstWouldAddEndPuncttrue
\mciteSetBstMidEndSepPunct{\mcitedefaultmidpunct}
{\mcitedefaultendpunct}{\mcitedefaultseppunct}\relax
\EndOfBibitem
\bibitem[Denisov \emph{et~al.}({2012})Denisov, Villanueva, Lorincz, May, and
  Wijngaarden]{Denisov2012}
D.~V. Denisov, Y.~Y. Villanueva, K.~A. Lorincz, S.~May and R.~J. Wijngaarden,
  \emph{{Phys. Rev. E}}, {2012}, \textbf{{85}}, 051309\relax
\mciteBstWouldAddEndPuncttrue
\mciteSetBstMidEndSepPunct{\mcitedefaultmidpunct}
{\mcitedefaultendpunct}{\mcitedefaultseppunct}\relax
\EndOfBibitem
\bibitem[Cleary(2008)]{Cleary2008}
P.~W. Cleary, \emph{Powder Technology}, 2008, \textbf{179}, 144\relax
\mciteBstWouldAddEndPuncttrue
\mciteSetBstMidEndSepPunct{\mcitedefaultmidpunct}
{\mcitedefaultendpunct}{\mcitedefaultseppunct}\relax
\EndOfBibitem
\bibitem[El~Shourbagy \emph{et~al.}({2006})El~Shourbagy, Morita, and
  Matuttis]{ElShourbagy2006}
S.~A.~M. El~Shourbagy, S.~Morita and H.-G. Matuttis, \emph{{J. Phys. Soc.
  Jpn.}}, {2006}, \textbf{{75}}, 104602\relax
\mciteBstWouldAddEndPuncttrue
\mciteSetBstMidEndSepPunct{\mcitedefaultmidpunct}
{\mcitedefaultendpunct}{\mcitedefaultseppunct}\relax
\EndOfBibitem
\bibitem[Campbell({2011})]{Campbell2011}
C.~S. Campbell, \emph{{Phys. Fluids}}, {2011}, \textbf{{23}}, 013306\relax
\mciteBstWouldAddEndPuncttrue
\mciteSetBstMidEndSepPunct{\mcitedefaultmidpunct}
{\mcitedefaultendpunct}{\mcitedefaultseppunct}\relax
\EndOfBibitem
\bibitem[{Anki Reddy} \emph{et~al.}({2009}){Anki Reddy}, Kumaran, and
  Talbot]{AnkiReddy2009}
K.~{Anki Reddy}, V.~Kumaran and J.~Talbot, \emph{{Phys. Rev. E}}, {2009},
  \textbf{{80}}, {031304}\relax
\mciteBstWouldAddEndPuncttrue
\mciteSetBstMidEndSepPunct{\mcitedefaultmidpunct}
{\mcitedefaultendpunct}{\mcitedefaultseppunct}\relax
\EndOfBibitem
\bibitem[{Anki Reddy} \emph{et~al.}({2010}){Anki Reddy}, Talbot, and
  Kumaran]{AnkiReddy2010}
K.~{Anki Reddy}, J.~Talbot and V.~Kumaran, \emph{{J. Fluid Mech.}}, {2010},
  \textbf{{660}}, {475}\relax
\mciteBstWouldAddEndPuncttrue
\mciteSetBstMidEndSepPunct{\mcitedefaultmidpunct}
{\mcitedefaultendpunct}{\mcitedefaultseppunct}\relax
\EndOfBibitem
\bibitem[B\"orzs\"onyi \emph{et~al.}(2012)B\"orzs\"onyi, Szab\'o, Wegner,
  Harth, T\"or\"ok, Somfai, Bien, and Stannarius]{Borzsonyi2012b}
T.~B\"orzs\"onyi, B.~Szab\'o, S.~Wegner, K.~Harth, J.~T\"or\"ok, E.~Somfai,
  T.~Bien and R.~Stannarius, \emph{Phys. Rev. E}, 2012, \textbf{86},
  051304\relax
\mciteBstWouldAddEndPuncttrue
\mciteSetBstMidEndSepPunct{\mcitedefaultmidpunct}
{\mcitedefaultendpunct}{\mcitedefaultseppunct}\relax
\EndOfBibitem
\bibitem[Skarp \emph{et~al.}(1979)Skarp, Lagerwall, Stebler, and
  McQueen]{Skarp1979}
K.~Skarp, S.~Lagerwall, B.~Stebler and D.~McQueen, \emph{Physica Scripta},
  1979, \textbf{19}, 339\relax
\mciteBstWouldAddEndPuncttrue
\mciteSetBstMidEndSepPunct{\mcitedefaultmidpunct}
{\mcitedefaultendpunct}{\mcitedefaultseppunct}\relax
\EndOfBibitem
\bibitem[G\"ahwiller(1972)]{Gaehwiller1972}
C.~G\"ahwiller, \emph{Phys. Rev. Lett.}, 1972, \textbf{28}, 1554\relax
\mciteBstWouldAddEndPuncttrue
\mciteSetBstMidEndSepPunct{\mcitedefaultmidpunct}
{\mcitedefaultendpunct}{\mcitedefaultseppunct}\relax
\EndOfBibitem
\bibitem[Beens and de~Jeu(1985)]{Beens1985}
W.~Beens and W.~de~Jeu, \emph{J. Chem. Phys.}, 1985, \textbf{82}, 3841\relax
\mciteBstWouldAddEndPuncttrue
\mciteSetBstMidEndSepPunct{\mcitedefaultmidpunct}
{\mcitedefaultendpunct}{\mcitedefaultseppunct}\relax
\EndOfBibitem
\bibitem[Ehrentraut and Hess(1995)]{Ehrentraut1995}
H.~Ehrentraut and S.~Hess, \emph{Phys. Rev. E}, 1995, \textbf{51}, 2203\relax
\mciteBstWouldAddEndPuncttrue
\mciteSetBstMidEndSepPunct{\mcitedefaultmidpunct}
{\mcitedefaultendpunct}{\mcitedefaultseppunct}\relax
\EndOfBibitem
\bibitem[Guo \emph{et~al.}(2012)Guo, Wassgren, Ketterhagen, Hancock, James, and
  Curtis]{Guo2012}
Y.~Guo, C.~Wassgren, W.~Ketterhagen, B.~Hancock, B.~James and J.~Curtis,
  \emph{J. Fluid Mech.}, 2012, \textbf{713}, 1\relax
\mciteBstWouldAddEndPuncttrue
\mciteSetBstMidEndSepPunct{\mcitedefaultmidpunct}
{\mcitedefaultendpunct}{\mcitedefaultseppunct}\relax
\EndOfBibitem
\bibitem[Desmond and Franklin(2006)]{Desmond2006}
K.~Desmond and S.~V. Franklin, \emph{Phys. Rev. E}, 2006, \textbf{73},
  031306\relax
\mciteBstWouldAddEndPuncttrue
\mciteSetBstMidEndSepPunct{\mcitedefaultmidpunct}
{\mcitedefaultendpunct}{\mcitedefaultseppunct}\relax
\EndOfBibitem
\bibitem[Kar\'atson \emph{et~al.}(2002)Kar\'atson, Sztan\'o, and
  Telbisz]{Karatson02}
D.~Kar\'atson, O.~Sztan\'o and T.~Telbisz, \emph{J. Sedimentary Res.}, 2002,
  \textbf{72}, 823\relax
\mciteBstWouldAddEndPuncttrue
\mciteSetBstMidEndSepPunct{\mcitedefaultmidpunct}
{\mcitedefaultendpunct}{\mcitedefaultseppunct}\relax
\EndOfBibitem
\bibitem[Elston and Smith(1970)]{Elston1970}
W.~E. Elston and E.~I. Smith, \emph{Geol. Soc. America Bull.}, 1970,
  \textbf{81}, 3393\relax
\mciteBstWouldAddEndPuncttrue
\mciteSetBstMidEndSepPunct{\mcitedefaultmidpunct}
{\mcitedefaultendpunct}{\mcitedefaultseppunct}\relax
\EndOfBibitem
\bibitem[Smith and Rhodes(1970)]{Smith1972}
E.~I. Smith and R.~C. Rhodes, \emph{Geol. Soc. America Bull.}, 1970,
  \textbf{83}, 1869\relax
\mciteBstWouldAddEndPuncttrue
\mciteSetBstMidEndSepPunct{\mcitedefaultmidpunct}
{\mcitedefaultendpunct}{\mcitedefaultseppunct}\relax
\EndOfBibitem
\bibitem[Rhodes and Smith(1972)]{Rhodes1972}
R.~C. Rhodes and E.~I. Smith, \emph{Geol. Soc. America Bull.}, 1972,
  \textbf{83}, 1863\relax
\mciteBstWouldAddEndPuncttrue
\mciteSetBstMidEndSepPunct{\mcitedefaultmidpunct}
{\mcitedefaultendpunct}{\mcitedefaultseppunct}\relax
\EndOfBibitem
\bibitem[Passchier({1987})]{Passchier1987}
C.~W. Passchier, \emph{{J. Struct. Geol.}}, {1987}, \textbf{{9}},
  {679--\&}\relax
\mciteBstWouldAddEndPuncttrue
\mciteSetBstMidEndSepPunct{\mcitedefaultmidpunct}
{\mcitedefaultendpunct}{\mcitedefaultseppunct}\relax
\EndOfBibitem
\bibitem[Passchier and Coelho({2006})]{Passchier2006}
C.~Passchier and S.~Coelho, \emph{Gondwana Research}, {2006}, \textbf{{10}},
  {66--76}\relax
\mciteBstWouldAddEndPuncttrue
\mciteSetBstMidEndSepPunct{\mcitedefaultmidpunct}
{\mcitedefaultendpunct}{\mcitedefaultseppunct}\relax
\EndOfBibitem
\bibitem[Mukherjee({2011})]{Mukherjee2011}
S.~Mukherjee, \emph{{Int. J. Earth Sci.}}, {2011}, \textbf{{100}},
  {1303--1314}\relax
\mciteBstWouldAddEndPuncttrue
\mciteSetBstMidEndSepPunct{\mcitedefaultmidpunct}
{\mcitedefaultendpunct}{\mcitedefaultseppunct}\relax
\EndOfBibitem
\bibitem[Passchier and Sokoutis({1993})]{Passchier1993}
C.~W. Passchier and D.~Sokoutis, \emph{J. Struct. Geol.}, {1993},
  \textbf{{15}}, {895--909}\relax
\mciteBstWouldAddEndPuncttrue
\mciteSetBstMidEndSepPunct{\mcitedefaultmidpunct}
{\mcitedefaultendpunct}{\mcitedefaultseppunct}\relax
\EndOfBibitem
\bibitem[ten Grotenhuis \emph{et~al.}({2002})ten Grotenhuis, Passchier, and
  Bons]{Grotenhuis2002}
S.~ten Grotenhuis, C.~Passchier and P.~Bons, \emph{{J. Struct. Geol.}}, {2002},
  \textbf{{24}}, {485--499}\relax
\mciteBstWouldAddEndPuncttrue
\mciteSetBstMidEndSepPunct{\mcitedefaultmidpunct}
{\mcitedefaultendpunct}{\mcitedefaultseppunct}\relax
\EndOfBibitem
\bibitem[ten Grotenhuis \emph{et~al.}({2003})ten Grotenhuis, Trouw, and
  Passchier]{Grotenhuis2003}
S.~ten Grotenhuis, R.~Trouw and C.~Passchier, \emph{{Tectonophysics}}, {2003},
  \textbf{{372}}, {1--21}\relax
\mciteBstWouldAddEndPuncttrue
\mciteSetBstMidEndSepPunct{\mcitedefaultmidpunct}
{\mcitedefaultendpunct}{\mcitedefaultseppunct}\relax
\EndOfBibitem
\bibitem[Oda \emph{et~al.}(1982)Oda, Konishi, and Nemat-Nasser]{Oda1982}
M.~Oda, J.~Konishi and S.~Nemat-Nasser, \emph{Mech. Materials}, 1982,
  \textbf{1}, 269\relax
\mciteBstWouldAddEndPuncttrue
\mciteSetBstMidEndSepPunct{\mcitedefaultmidpunct}
{\mcitedefaultendpunct}{\mcitedefaultseppunct}\relax
\EndOfBibitem
\bibitem[Oda \emph{et~al.}(1985)Oda, Nemat-Nasser, and Konichi]{Oda1985}
M.~Oda, S.~Nemat-Nasser and J.~Konichi, \emph{Soil Found.}, 1985, \textbf{25},
  85\relax
\mciteBstWouldAddEndPuncttrue
\mciteSetBstMidEndSepPunct{\mcitedefaultmidpunct}
{\mcitedefaultendpunct}{\mcitedefaultseppunct}\relax
\EndOfBibitem
\bibitem[Shodja and Nezami({2003})]{Shodja2003}
H.~Shodja and E.~Nezami, \emph{Int. J. Numer. Anal. Methods in Geomech.},
  {2003}, \textbf{{27}}, {403}\relax
\mciteBstWouldAddEndPuncttrue
\mciteSetBstMidEndSepPunct{\mcitedefaultmidpunct}
{\mcitedefaultendpunct}{\mcitedefaultseppunct}\relax
\EndOfBibitem
\bibitem[Pe\~na \emph{et~al.}({2007})Pe\~na, Garcia-Rojo, and Herrmann]{Pena2007}
A.~A. Pe\~na, R.~Garcia-Rojo and H.~J. Herrmann, \emph{{Granular Matter}},
  {2007}, \textbf{{9}}, {279--291}\relax
\mciteBstWouldAddEndPuncttrue
\mciteSetBstMidEndSepPunct{\mcitedefaultmidpunct}
{\mcitedefaultendpunct}{\mcitedefaultseppunct}\relax
\EndOfBibitem
\bibitem[Atwell and Olafsen({2005})]{Atwell2005}
J.~Atwell and J.~S. Olafsen, \emph{{Phys. Rev. E}}, {2005}, \textbf{{71}},
  {062301}\relax
\mciteBstWouldAddEndPuncttrue
\mciteSetBstMidEndSepPunct{\mcitedefaultmidpunct}
{\mcitedefaultendpunct}{\mcitedefaultseppunct}\relax
\EndOfBibitem
\bibitem[Harth \emph{et~al.}(2011)Harth, H\"ome, Trittel, Kornek, Strachauer,
  and Will]{Harth2011}
K.~Harth, S.~H\"ome, T.~Trittel, U.~Kornek, U.~Strachauer and K.~Will,
  \emph{Proc. 20th ESA Symposium Hy\'eres}, 2011,  p. 493\relax
\mciteBstWouldAddEndPuncttrue
\mciteSetBstMidEndSepPunct{\mcitedefaultmidpunct}
{\mcitedefaultendpunct}{\mcitedefaultseppunct}\relax
\EndOfBibitem
\bibitem[Harth \emph{et~al.}(2013)Harth, Kornek, Trittel, Strachauer, H\"ome,
  Will, and Stannarius]{Harth2013}
K.~Harth, U.~Kornek, T.~Trittel, U.~Strachauer, S.~H\"ome, K.~Will and
  R.~Stannarius, \emph{Phys. Rev. Lett.}, 2013, \textbf{110}, 144102\relax
\mciteBstWouldAddEndPuncttrue
\mciteSetBstMidEndSepPunct{\mcitedefaultmidpunct}
{\mcitedefaultendpunct}{\mcitedefaultseppunct}\relax
\EndOfBibitem
\bibitem[Daniels \emph{et~al.}({2009})Daniels, Park, Lubensky, and
  Durian]{Daniels2009}
L.~J. Daniels, Y.~Park, T.~C. Lubensky and D.~J. Durian, \emph{{Phys. Rev. E}},
  {2009}, \textbf{{79}}, {041301}\relax
\mciteBstWouldAddEndPuncttrue
\mciteSetBstMidEndSepPunct{\mcitedefaultmidpunct}
{\mcitedefaultendpunct}{\mcitedefaultseppunct}\relax
\EndOfBibitem
\bibitem[Daniels and Durian({2011})]{Daniels2011}
L.~J. Daniels and D.~J. Durian, \emph{{Phys. Rev. E}}, {2011}, \textbf{{83}},
  {061304}\relax
\mciteBstWouldAddEndPuncttrue
\mciteSetBstMidEndSepPunct{\mcitedefaultmidpunct}
{\mcitedefaultendpunct}{\mcitedefaultseppunct}\relax
\EndOfBibitem
\bibitem[Viot and Talbot({2004})]{Viot2004}
P.~Viot and J.~Talbot, \emph{{Phys. Rev. E}}, {2004}, \textbf{{69}},
  {051106}\relax
\mciteBstWouldAddEndPuncttrue
\mciteSetBstMidEndSepPunct{\mcitedefaultmidpunct}
{\mcitedefaultendpunct}{\mcitedefaultseppunct}\relax
\EndOfBibitem
\bibitem[Gomart \emph{et~al.}({2005})Gomart, Talbot, and Viot]{Gomart2005}
H.~Gomart, J.~Talbot and P.~Viot, \emph{{Phys. Rev. E}}, {2005}, \textbf{{71}},
  {051306}\relax
\mciteBstWouldAddEndPuncttrue
\mciteSetBstMidEndSepPunct{\mcitedefaultmidpunct}
{\mcitedefaultendpunct}{\mcitedefaultseppunct}\relax
\EndOfBibitem
\bibitem[Piasecki and Viot({2006})]{Piasecki2006}
J.~Piasecki and P.~Viot, \emph{{Europhys. Lett.}}, {2006}, \textbf{{74}},
  {1--7}\relax
\mciteBstWouldAddEndPuncttrue
\mciteSetBstMidEndSepPunct{\mcitedefaultmidpunct}
{\mcitedefaultendpunct}{\mcitedefaultseppunct}\relax
\EndOfBibitem
\bibitem[Piasecki \emph{et~al.}({2007})Piasecki, Talbot, and
  Viot]{Piasecki2007}
J.~Piasecki, J.~Talbot and P.~Viot, \emph{{Phys. Rev. E}}, {2007},
  \textbf{{75}}, {051307}\relax
\mciteBstWouldAddEndPuncttrue
\mciteSetBstMidEndSepPunct{\mcitedefaultmidpunct}
{\mcitedefaultendpunct}{\mcitedefaultseppunct}\relax
\EndOfBibitem
\bibitem[Chrzanowska and Ehrentraut(2002)]{Chrzanowska2002}
A.~Chrzanowska and H.~Ehrentraut, \emph{Technische Mechanik}, 2002,
  \textbf{22}, 56\relax
\mciteBstWouldAddEndPuncttrue
\mciteSetBstMidEndSepPunct{\mcitedefaultmidpunct}
{\mcitedefaultendpunct}{\mcitedefaultseppunct}\relax
\EndOfBibitem
\bibitem[Chrzanowska and Ehrentraut(2003)]{Chrzanowska2003-1}
A.~Chrzanowska and H.~Ehrentraut, in \emph{Dynamic response of granular and
  porous materials under large and catastrophic deformations}, ed. K.~Hutter
  and N.~P. Kirchner, Springer, Berlin, 2003, p. 315\relax
\mciteBstWouldAddEndPuncttrue
\mciteSetBstMidEndSepPunct{\mcitedefaultmidpunct}
{\mcitedefaultendpunct}{\mcitedefaultseppunct}\relax
\EndOfBibitem
\bibitem[Foulaadvand and Yarifard({2011})]{Foulaadvand2011}
M.~Foulaadvand and M.~Yarifard, \emph{{Eur. Phys. J. E}}, {2011},
  \textbf{{34}}, {41}\relax
\mciteBstWouldAddEndPuncttrue
\mciteSetBstMidEndSepPunct{\mcitedefaultmidpunct}
{\mcitedefaultendpunct}{\mcitedefaultseppunct}\relax
\EndOfBibitem
\bibitem[Frenkel(1987)]{Frenkel1987}
D.~Frenkel, \emph{Mol. Cryst.}, 1987, \textbf{60}, 1\relax
\mciteBstWouldAddEndPuncttrue
\mciteSetBstMidEndSepPunct{\mcitedefaultmidpunct}
{\mcitedefaultendpunct}{\mcitedefaultseppunct}\relax
\EndOfBibitem
\bibitem[Kr\"oger(2004)]{Kroger2004}
J.~Kr\"oger, \emph{Phys. Rep.}, 2004, \textbf{390}, 453\relax
\mciteBstWouldAddEndPuncttrue
\mciteSetBstMidEndSepPunct{\mcitedefaultmidpunct}
{\mcitedefaultendpunct}{\mcitedefaultseppunct}\relax
\EndOfBibitem
\bibitem[Wildman \emph{et~al.}({2009})Wildman, Beecham, and
  Freeman]{Wildman2009}
R.~Wildman, J.~Beecham and T.~Freeman, \emph{{Eur. Phys. J. Special Topics}},
  {2009}, \textbf{{179}}, {5}\relax
\mciteBstWouldAddEndPuncttrue
\mciteSetBstMidEndSepPunct{\mcitedefaultmidpunct}
{\mcitedefaultendpunct}{\mcitedefaultseppunct}\relax
\EndOfBibitem
\bibitem[Costantini \emph{et~al.}({2005})Costantini, Marconi, Kalibaeva, and
  Ciccotti]{Costantini2005}
G.~Costantini, U.~Marconi, G.~Kalibaeva and G.~Ciccotti, \emph{{J. Chem.
  Phys.}}, {2005}, \textbf{{122}}, 164505\relax
\mciteBstWouldAddEndPuncttrue
\mciteSetBstMidEndSepPunct{\mcitedefaultmidpunct}
{\mcitedefaultendpunct}{\mcitedefaultseppunct}\relax
\EndOfBibitem
\bibitem[Kanzaki \emph{et~al.}({2010})Kanzaki, {Cruz Hidalgo}, Maza, and
  Pagonabarraga]{Kanzaki2010}
T.~Kanzaki, R.~{Cruz Hidalgo}, D.~Maza and I.~Pagonabarraga, \emph{{J. Stat.
  Mech. - Theor. Exp.}}, {2010},  P06020\relax
\mciteBstWouldAddEndPuncttrue
\mciteSetBstMidEndSepPunct{\mcitedefaultmidpunct}
{\mcitedefaultendpunct}{\mcitedefaultseppunct}\relax
\EndOfBibitem
\bibitem[Huthmann \emph{et~al.}({1999})Huthmann, Aspelmeier, and
  Zippelius]{Huthmann1999}
M.~Huthmann, T.~Aspelmeier and A.~Zippelius, \emph{{Phys. Rev. E}}, {1999},
  \textbf{{60}}, {654}\relax
\mciteBstWouldAddEndPuncttrue
\mciteSetBstMidEndSepPunct{\mcitedefaultmidpunct}
{\mcitedefaultendpunct}{\mcitedefaultseppunct}\relax
\EndOfBibitem
\bibitem[Villemot and Talbot({2012})]{Villemot2012}
F.~Villemot and J.~Talbot, \emph{{Granular Matter}}, {2012}, \textbf{{14}},
  {91}\relax
\mciteBstWouldAddEndPuncttrue
\mciteSetBstMidEndSepPunct{\mcitedefaultmidpunct}
{\mcitedefaultendpunct}{\mcitedefaultseppunct}\relax
\EndOfBibitem
\bibitem[Meyer \emph{et~al.}(1975)Meyer, Li\'ebert, Strelecki, and
  Keller]{Meyer1975}
R.~B. Meyer, L.~Li\'ebert, L.~Strelecki and P.~Keller, \emph{J. Physique
  (Lett.)}, 1975, \textbf{36}, L--69\relax
\mciteBstWouldAddEndPuncttrue
\mciteSetBstMidEndSepPunct{\mcitedefaultmidpunct}
{\mcitedefaultendpunct}{\mcitedefaultseppunct}\relax
\EndOfBibitem
\bibitem[Takezoe and Takanishi(2006)]{Takezoe2006}
H.~Takezoe and Y.~Takanishi, \emph{Jpn. J. Appl. Phys.}, 2006, \textbf{45},
  597\relax
\mciteBstWouldAddEndPuncttrue
\mciteSetBstMidEndSepPunct{\mcitedefaultmidpunct}
{\mcitedefaultendpunct}{\mcitedefaultseppunct}\relax
\EndOfBibitem
\bibitem[Carlsson and Skarp(1986)]{Carlsson1986}
T.~Carlsson and K.~Skarp, \emph{Liq. Cryst.}, 1986, \textbf{5}, 455\relax
\mciteBstWouldAddEndPuncttrue
\mciteSetBstMidEndSepPunct{\mcitedefaultmidpunct}
{\mcitedefaultendpunct}{\mcitedefaultseppunct}\relax
\EndOfBibitem
\end{mcitethebibliography}
\end{document}